\documentclass[arguments]{aastex631}

\shorttitle{spectral property of high linearly polarized signal from PSR J0332+5434}
\shortauthors{Mitra, Basu \& Melikidze}

\begin{document}

\title{On the flux density spectral property of high linearly polarized signal from Pulsar J0332+5434}

\author[0000-0002-9142-9835]{Dipanjan Mitra}
\affiliation{National Centre for Radio Astrophysics, Tata Institute of Fundamental Research, Pune 411007, India.}
\affiliation{Janusz Gil Institute of Astronomy, University of Zielona G\'ora, ul. Szafrana 2, 65-516 Zielona G\'ora, Poland.}

\author[0000-0003-1824-4487]{Rahul Basu}
\affiliation{Janusz Gil Institute of Astronomy, University of Zielona G\'ora, ul. Szafrana 2, 65-516 Zielona G\'ora, Poland.}

\author[0000-0003-1879-1659]{George I. Melikidze}
\affiliation{Janusz Gil Institute of Astronomy, University of Zielona G\'ora, ul. Szafrana 2, 65-516 Zielona G\'ora, Poland.}
\affiliation{Evgeni Kharadze Georgian National Astrophysical Observatory, 0301 Abastumani, Georgia.}

\begin{abstract}
The polarization position angles (PPA) of time samples with high  
linear polarization often show two parallel tracks across the pulsar profile 
that follow the rotating vector model (RVM). This feature 
support coherent curvature radiation (CCR) as the 
underlying mechanism of radio emission from pulsars, where the parallel tracks 
of the PPA represent the orthogonal extraordinary X and ordinary O eigen modes of strongly 
magnetized pair plasma. However, the frequency evolution of these high linearly
polarized signals remains unexplored. In this work we explore the flux density spectral 
nature of high linearly polarized signals by studying the emission from PSR 
J0332+5434 over a frequency range between 300 MHz and 750 MHz,
using the Giant Metrewave Radio Telescope. The pulsar average profile comprises
of a central core and a pair of conal components. We find the high linearly 
polarized time samples to be broadband in nature and in many cases they 
resemble a narrow spiky feature in the conal regions. These spiky features are 
localised within a narrow pulse longitude, over the entire frequency range,
and their spectral shapes sometimes resemble an inverted parabolic shape. In all such 
cases the PPA are exclusively along one of the orthogonal RVM tracks, likely 
corresponding to the X-mode. The inverted spectral shape can in principle be 
explained if the high linearly polarized emission in these time samples are
formed due to incoherent addition of CCR from a large number of charged 
solitons (charge bunches) exciting the X-mode.
\end{abstract}


\section{Introduction} \label{sec:intro}

PSR J0332+5434 (B0329+54) discovered in 1968 \citep{1968Natur.219..574C} is one
of the brightest pulsars in the northern sky and as a result has been 
extensively studied over the radio spectrum from about 25 MHz to 43 GHz. Over a
majority of this frequency range the pulsar profiles have three distinct 
emission components, consisting of a dominant central core surrounded by a pair
of leading and trailing conal components, but the profiles are scattered by the
inter stellar medium (ISM) below 100 MHz. The core component has a composite 
nature and becomes comparatively weaker with increasing frequency \citep[see 
e.g.][]{2017MNRAS.468.4389S,2021ApJ...917...48B}. The profile type has been 
classified as core-outer cone triple \citep{1993ApJ...405..285R,
1993ApJS...85..145R}, although some studies also reported the presence of 
additional inner cones \citep{2001ApJ...555...31G}. The pulsar profile shows 
the effect of radius to frequency mapping where the profile width progressively
decreases at higher frequencies \citep{2002ApJ...577..322M}. The pulsar 
exhibits moderate levels of average linear and circular polarization across 
frequencies. The average polarization position angle (PPA) has a complex 
traverse across the profile, but single pulse observations have revealed two 
parallel PPA tracks \citep{1995MNRAS.276L..55G,2007MNRAS.379..932M} that follow
the rotating vector model \citep[][hereafter RVM]{1969ApL.....3..225R}. 
According to RVM the linearly polarized emission traces the change in the 
dipolar magnetic field line plane of the rotating pulsar, and the model has 
been extensively used to measure the location of the radio emission region and 
emission geometry in several pulsars \citep[see][for a 
review]{2017JApA...38...52M,2024Univ...10..248M}.

The average flux density spectra in pulsars generally exhibit a steep power law
nature ($S\propto\nu^{\alpha}$) above 100 MHz with typical spectral indices, 
$\alpha\sim-1.8$ \citep{2000A&AS..147..195M,2018MNRAS.473.4436J}. However, the 
spectrum of PSR J0332+5434 deviates from this typical behaviour and can be 
approximated as a broken power law, where at the high frequency range above 1 
GHz the spectrum is steeper, while between 100 MHz and 1 GHz it becomes much 
flatter. The flux densities in pulsars usually show large fluctuations due to 
variations in the parameter of emission mechanism as well as scintillations 
during propagation in the ISM. As a result the spectral measurement requires 
averaging the flux density over longer timescales, usually several thousand 
pulses, to obtain a stable value. One of the drawbacks of the flux density 
estimates is that in the averaging process the information of the individual 
emission sources in the pulsar plasma, which last for significantly shorter 
timescales, is wiped out. Hence, our understanding of the coherent radio 
emission mechanism produced by these individual emission can benefit from the 
study of single pulse spectra across frequencies. The initial attempt at 
characterising the single pulse spectra was conducted by \citet[][hereafter 
K03]{2003A&A...407..655K}, who carried out simultaneous observations of two 
pulsars, J0332+5434 and J1136+1551, over a wide frequency range between 200 MHz
and 5 GHz. Their study showed that the single pulse flux densities have 
stronger correlation between nearby frequencies and the correlation gets weaker
when widely separated frequencies are considered. This behaviour indicates that 
the radiation source in pulsars influences the single pulse spectra and 
motivates this present work of using single pulse spectral properties to study 
the underlying emission process. 

In PSR J0332+5434, K03 found the spectra of the single pulses to be similar to 
the average spectrum, suggesting the possibility of a relatively stable and 
unvarying emission process on shorter timescales, although this does not 
provide any further constraints on the underlying emission mechanism. Recent 
studies have uncovered that PPAs of high linearly polarized time samples in 
single pulses are oriented along two parallel tracks, separated by 90$\degr$~in
phase, that closely follow the RVM, even in pulsars where the average PPA shows
complex non-RVM like PPA traverse \citep{2023MNRAS.521L..34M,2023ApJ...952..151M,
2024MNRAS.530.4839J}. 
The feature that the high linearly polarized time sample follow 
the RVM can be only explained by curvature radiation, 
since the waves excited by curvature radiation are polarized 
parallel or perpendicular to the curved magnetic field line planes (see \citealt{2024Univ...10..248M}
for a recent review).
Further the requirement of high brightness temperature
of the pulsar radio emission establishes coherent curvature radiation (CCR) from 
charged bunches as the most likely emission mechanism. The high linearly 
polarized signals are free from depolarization and correspond exclusively to a 
single polarization state. The two orthogonal polarization states of the PPA 
can be associated with the extraordinary (X) and ordinary (O) eigen modes of a 
strongly magnetized pair plasma excited by CCR, with the polarization vector 
oriented perpendicular and parallel to the magnetic field line planes, 
respectively. Hence, it is expected that the spectra of these high linearly 
polarized time samples should provide additional constraints on the emission 
mechanism by bearing resemblance to the CCR spectra of individual charged 
bunches.

In this paper we explore the spectral nature of high linearly polarized time 
samples by carrying out simultaneous observations of single pulse polarization 
from PSR J0332+5434 over a frequency range, between 300 MHz and 750 MHz, 
using the Giant Meterwave Radio Telescope (GMRT,\citealt{1991CuSc...60...95S}). We intend to examine whether 
these highly polarized time samples are broadband in nature and find their 
spectral behaviour over the observed frequency range. In section~\ref{sec2} we 
describe the observations and data analysis methods. Section~\ref{sec3} reports 
the spectral properties of the high linearly polarized time samples while 
section~\ref{sec4} provides a qualitative understanding of the spectra from the 
perspective of CCR from charge bunches. In section~\ref{sec5} we summarize the 
spectral nature of the highly polarized time samples of PSR J0332+5434.

\section{Observation \& Analysis} \label{sec2}

A group of bright pulsars, including PSR J0332+5434, were observed using the 
GMRT between December 2019 and February 2020. The 
GMRT is an interferometer consisting of 30 antennas, each of 45 meter diameter,
and upgraded wideband receiver systems \citep{2017CSci..113..707G} operating 
at four distinct frequency bands between 120-250 MHz (Band2), 250-500 MHz 
(Band-3), 550-850 MHz (Band-4), and 1050-1450 MHz (Band-5). The antennas are 
arranged resembling a Y-shaped array with two distinct configurations, a 
central square populated by 14 antennas spread within an area of 1 square km, 
and 16 antennas along three arms within a circle of 25 km diameter. The high 
time resolution observations for studying pulsars usually employs the 
phased-array mode where the signals from different antennas are co-added in 
phase. However, accurate flux density measurements require interferometric 
imaging. The GMRT allows the simultaneous measurements in both phased-array and
interferometric modes.

The whole array can be divided into several subarrays, with different sets of 
antennas, observing simultaneously at multiple frequency bands. PSR J0332+5434 
was observed on 17 January 2020 using dual subarrays at Band-3 and Band-4, with
200 MHz bandwidth, such that the pulsar emission could be recorded with a near 
continuous frequency coverage between 300 MHz and 750 MHz. During this session 
a total of 26 antennas were available for observations, with 3 antennas from 
the central square and 1 from an arm affected by technical issues. The first 
subarray in Band-3 (300-500 MHz) had 9 antennas, comprising of 6 central square
antennas and 3 arm antennas, while the remaining 17 antennas were observing in 
Band-4 (550-750 MHz), comprising of 5 central square antennas and 12 arm 
antennas. The phased array used the central square antennas and the first two 
arm antennas while all available antennas in each subarray were part of the 
interferometric measurements.  More central square antennas were part of 
the Band-3 subarray as the de-phasing of the nearby antennas would be slower 
allowing longer scans for single pulse studies.
All four polarizations in the full Stokes mode were measured during these 
observations with the 200 MHz band divided into 2048 spectral channels. The 
entire 400 MHz frequency range was further subdivided into 6 subbands of 50-60 
MHz (see Table~\ref{tab:flux}), after removing channels affected by Radio 
frequency Interference (RFI), to study the spectral nature of the emission. The
output signals from the phased array, that can be considered equivalent to a 
single dish, were sampled at high temporal resolution of 327.68 microseconds. 
The interferometer on the other hand recorded visibilities corresponding to 
each antenna pair in the subarray with 2 seconds integration time.

\subsection{Estimating Flux Density}
\begin{figure}
\epsscale{1.00}
\gridline{\fig{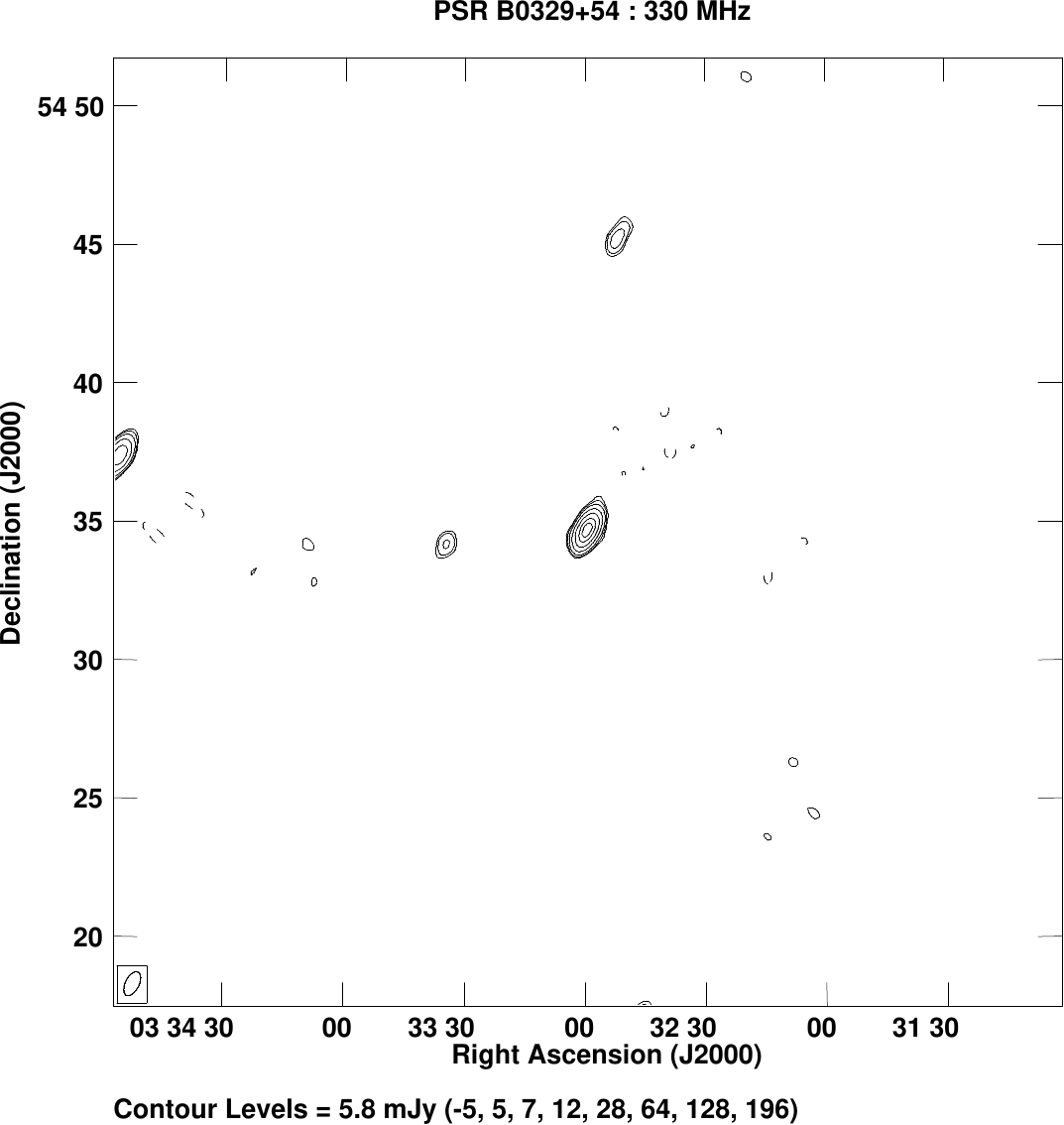}{0.38\textwidth}{}
          \fig{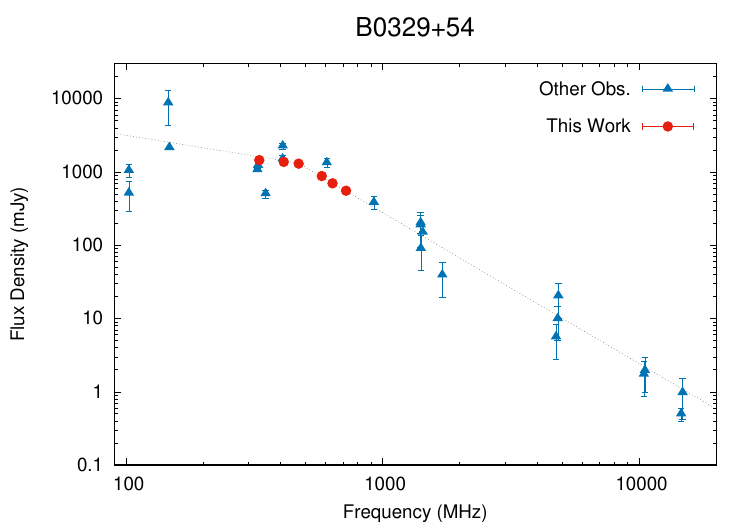}{0.59\textwidth}{}
	 }
\caption{The left panel shows the total intensity contour map of the region of 
sky around PSR J0332+5434 at 330 MHz. The pulsar is seen as an unresolved point 
source at the center of the image, along with a few other sources in the field
of view. The spatial resolution of the image is shown as a synthesized beam in
the bottom left corner of the image. The right panel shows the spectral 
behaviour of PSR J0332+5434 with flux density measurements from the literature,
between 100 MHz and 20 GHz (blue triangle), compiled by 
\cite{2022PASA...39...56S}, as well as the 6 measurements between 300 MHz and 
750 MHz (red circle) reported in Table.~\ref{tab:flux}. The pulsar shows a 
broken power law spectrum with a transition from a steeper spectral index to a 
more flatter behaviour around 470 MHz.
\label{fig:flux}}
\end{figure}

The measurement of the average flux density required observations of a standard
flux calibrator, 3C48, whose flux levels at these observing frequencies have 
been well established \citep{PB17}. In addition, a bright point like source, 
0432+416, located close to the pulsar, was observed for 3-4 minutes before and 
after the pulsar emission was recorded to correct for amplitude and phase 
variations in antenna responses with time as well as across the frequency band.
The pulsar was observed for 85 minutes resulting in around 7000 single pulses. 
The maps of each of the 6 subbands were created using the Astronomical Image 
Processing System (AIPS) software. Fig \ref{fig:flux}, left panel, shows an 
image of the region of sky around PSR J0332+5434 at 330 MHz, as a total 
intensity contour map. The pulsar is seen as an unresolved point source at the 
center of the image. The unequal number of antennas in the two subarrays, 
particularly the use of more central square antennas in Band-3, resulted in the
the noise rms in the maps to vary and the spatial resolutions being much larger
in case of the lower frequency measurement. However, in all cases the pulsar 
was clearly seen in the images as a point source with high detection 
sensitivity. The average flux densities of PSR J0332+5434 at the 6 frequencies
are reported in Table \ref{tab:flux} along with the noise rms in the maps, 
their spatial resolutions, the flux density of the flux calibrator (3C48) and 
the phase calibrator (0432+416) used to scale the source flux density. In the 
right panel of Fig \ref{fig:flux} the pulsar spectrum between 100 MHz and 20 
GHz is shown, where the measurements at different frequencies have been 
compiled by \citet{2022PASA...39...56S}. The measurements carried out in this 
work are also shown in the figure and appear to be consistent with earlier 
reported values in the relevant frequency range. The pulsar has a broken power 
law spectrum with a steeper spectral index at higher frequencies around 1 GHz 
and above and a much flatter spectra at low frequencies around 100 MHz. Our 
wideband observations provide a more accurate estimate of the transition 
frequency between the two spectral behaviour to be around 470 MHz. The spectral
indices obtained from the fits (black dotted lines) in Fig \ref{fig:flux} 
(right panel) are $\alpha_{low}=-0.5\pm0.1$ and $\alpha_{high}=-2.06\pm0.07$. 

\begin{deluxetable*}{ccccccc}
\tablecaption{Estimating Flux density of PSR J0332+5434\label{tab:flux}}
\tablewidth{0pt}
\tablehead{
 \colhead{Frequency} & \colhead{Bandwidth} & \colhead{Image rms} & \colhead{Resolution} & \colhead{3C48$^*$} & \colhead{0432+416} & \colhead{J0332+5434} \\
 \colhead{(MHz)} & \colhead{(MHz)} & \colhead{(mJy)} & \colhead{($"\times"$)} & \colhead{(Jy)} & \colhead{(Jy)} & \colhead{(Jy)}}
\startdata
  330.6 & 55.7 & 8.2 & 57.1$\times$29.4 & 43.51 & 15.33$\pm$0.44 & 1.447$\pm$0.044 \\
  412.0 & 58.5 & 4.4 & 40.6$\times$23.7 & 38.45 & 15.33$\pm$0.42 & 1.372$\pm$0.039 \\
  470.7 & 50.8 & 5.4 & 43.8$\times$20.8 & 35.50 & 14.74$\pm$0.41 & 1.298$\pm$0.038 \\
  579.3 & 54.7 & 0.89 & 5.9$\times$3.2 & 31.11 & 14.60$\pm$0.39 & 0.877$\pm$0.024 \\
  637.9 & 58.5 & 0.75 & 5.3$\times$3.2 & 29.18 & 13.98$\pm$0.38 & 0.697$\pm$0.019 \\
  720.5 & 58.5 & 0.71 & 5.0$\times$2.9 & 26.84 & 13.15$\pm$0.36 & 0.554$\pm$0.015 \\
\enddata
\tablecomments{$^*$The flux scale is estimated from \citet{PB17}.}
\end{deluxetable*}

\begin{figure}
\gridline{\fig{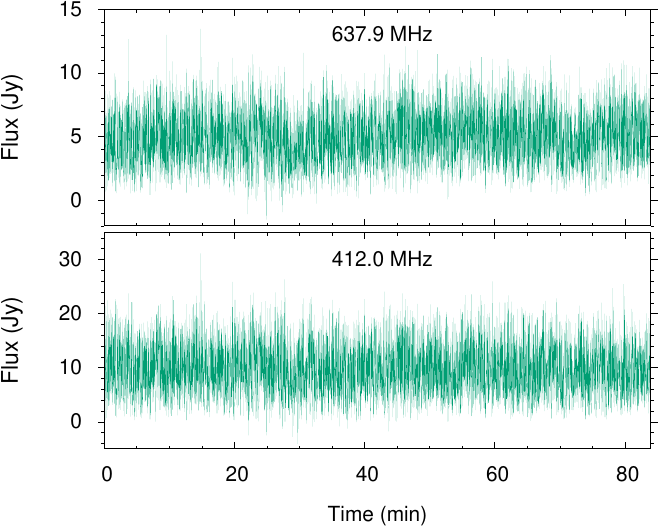}{0.52\textwidth}{}
          \fig{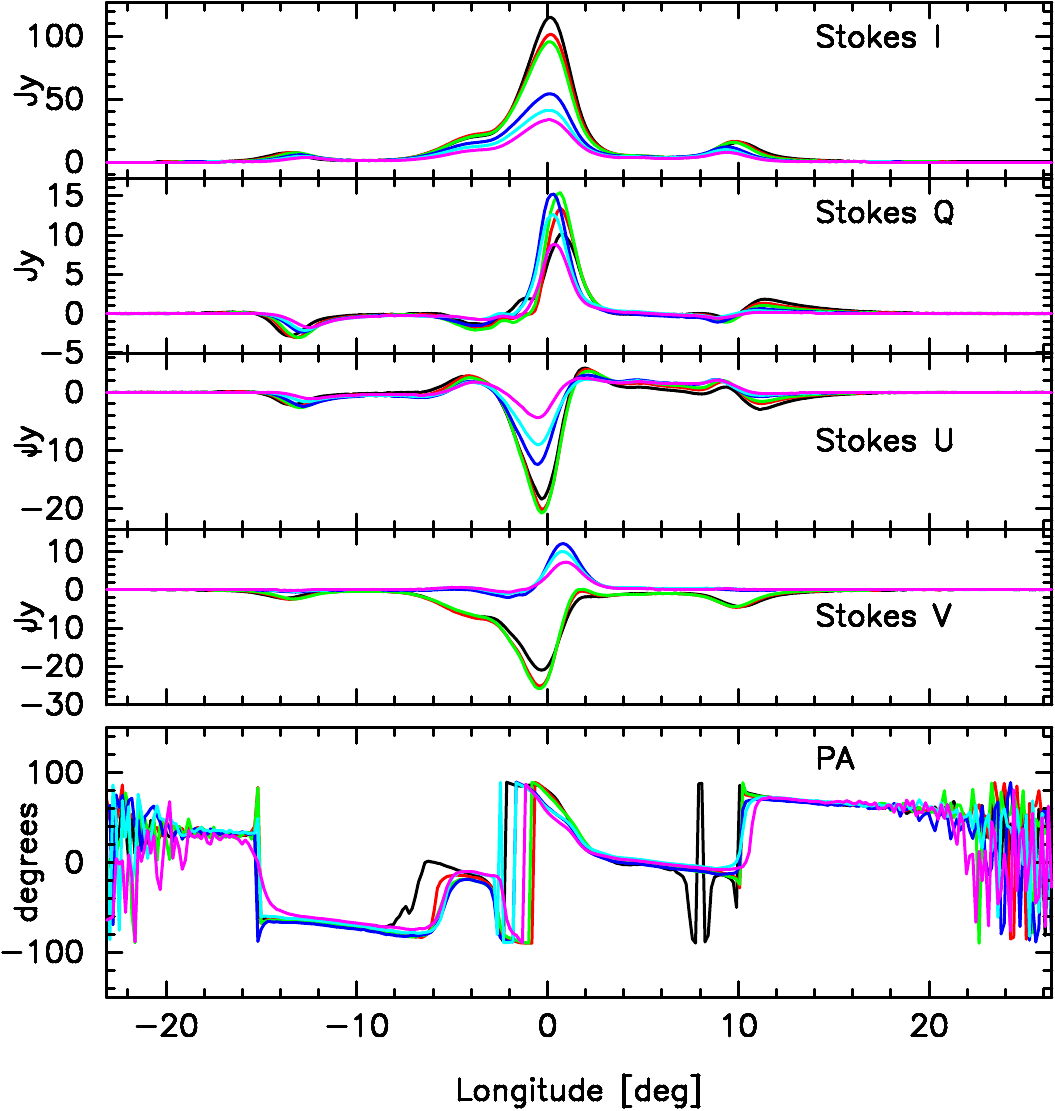}{0.42\textwidth}{}
         }
\caption{The figure shows the pulsed emission from PSR J0332+5434 after 
applying the flux scaling from the image analysis. The left panel shows the
average energy within the pulse window in each single pulse as a function of 
time for the two frequencies, 637.9 MHz (top window) and 412.0 MHz (bottom 
window). The right panel shows the average profile at all 6 frequencies, 
330.6 MHz (black), 412.0 MHz (red), 470.7 MHz (green), 579.3 MHz (blue), 637.9 
MHz (cyan) and 720.5 MHz (magneta). The four Stokes parameters I, Q, U, V are
shown along with the polarization position angle across the pulse window.
\label{fig:fluxpuls}}
\end{figure}

The phased-array measurements are in arbitrary units that can be converted into
the unit of flux density (Jy) with proper scaling from the interferometric 
estimates. The average flux densities in each subband obtained from the images 
(see Table \ref{tab:flux}) correspond to the average counts in the total 
intensity profiles, i.e. the emission in the pulse window equally distributed
across the entire period. Additionally, the profiles have arbitrary baseline
levels that needs to be subtracted before finding the period averaged counts. 
The scaling factor is estimated for each subband by dividing the measured flux 
density with baseline subtracted average counts in the profile. The intensity 
levels of the single pulses for all four Stokes parameters (I, Q, U and V) were
multiplied by the scaling factor to convert the arbitrary units into the 
emitted flux density from the pulsar. Fig~\ref{fig:fluxpuls}, left panel, shows
the results of flux scaling on the total intensity single pulse emission, where
the average intensity in the pulsed window is represented as function of 
observing time at two frequencies, 412.0 MHz and 637.9 MHz. The right panel of 
the figure shows the flux calibrated average profile of the four Stokes 
parameters as well as the variation of the PPA across the profile for the six 
frequency subbands. 

One of the primary issues affecting pulsar flux measurements and proper 
estimations of the spectral nature is interstellar scintillations 
\citep{1990ARA&A..28..561R,1998ApJ...507..846C}. The scintillations can be 
divided into diffractive and refractive classes causing intensity modulations 
over various timescales as well as frequency ranges. However, K03 showed that
in PSR J0332+5434 the effect of scintillations at frequencies below 1 GHz is 
not prominent. This is also corroborated in our analysis, where the measured 
average flux densities (Fig \ref{fig:flux}, right panel), is consistent with 
previous studies. Further, there is no sign of any periodic/quasi-periodic 
modulations of the single pulse intensities over the observing duration in any
of the frequency bands (see left panel in Fig \ref{fig:fluxpuls}). We conclude
that any spectral feature seen in the single pulse emission can be attributed
to the intrinsic emission mechanism. 

\begin{figure}
\plottwo{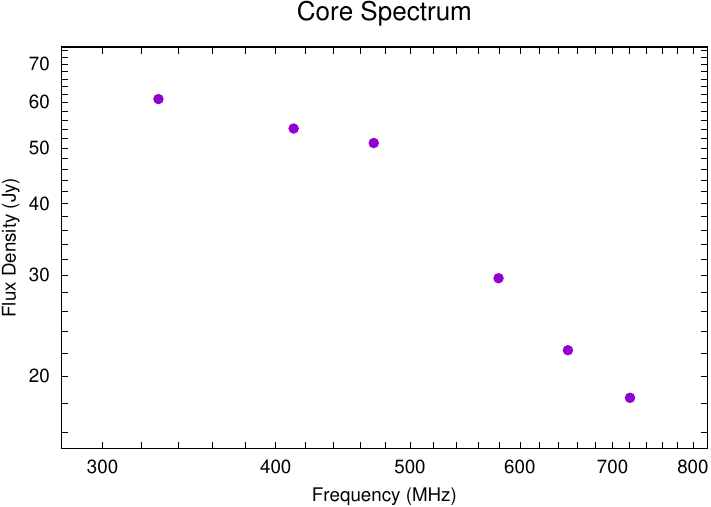}{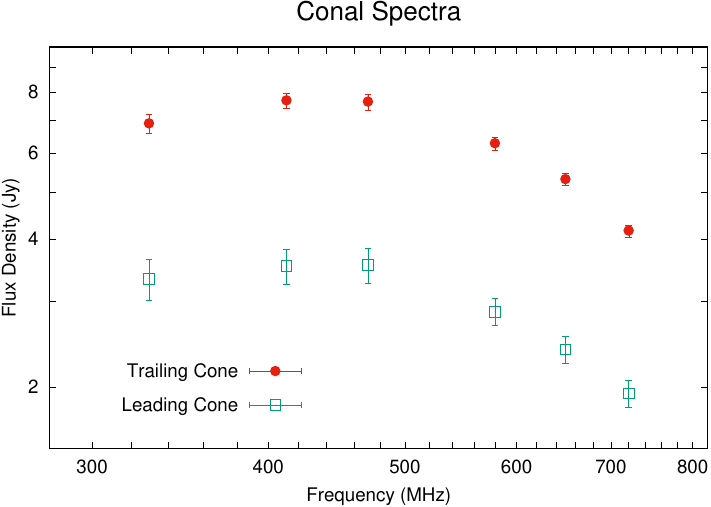}
\caption{The left plot shows the average Flux density of the core component 
measured at the 6 frequency bands between 300 MHz and 750 MHz. The right panel
shows the corresponding measurements for the leading and trailing conal 
components. All three components show the transition from the steep spectral
behaviour above 470 MHz, to a much flatter spectra below it. The core component
has a power law spectral behaviour at the lower frequency range, while the 
conal components show flat spectra with slight turnover.
\label{fig:avgcompspect}}
\end{figure}

The profile of PSR J0332+5434 consists of three main components, a pair of 
leading and trailing cones as well as a prominent composite central core, that 
are well separated from each other. We have estimated the average flux densities
of all three components at the 6 subbands. The component windows used for the 
flux estimation were selected to be between 10\% level of the peak intensities.
Fig \ref{fig:avgcompspect} shows the spectral nature of the three components 
between 300 MHz and 750 MHz, with all three components shifting from a steep 
power law spectra to a much flatter behaviour around 470 MHz. The core 
component maintains a negative power law spectrum below 470 MHz, but the conal 
components show slight turnover in their spectral shape with a positive 
spectral index.

The high intensity level of the pulsar in certain instances saturated the 
telescope response, where the peak intensities of the core component in the 
single pulses exceeded the detection limit and were suppressed. One clear 
indication of saturation was that the linear polarization level (L = 
$\sqrt{\rm{Q}^2+\rm{U}^2}$) in these time samples was much higher than the 
total intensity. We carefully examined the entire pulse sequence and found 
saturation affecting less then 10\% of the single pulses in each subband and 
limited to less than 5\% of the emission window. It is unlikely that the flux 
estimates from the images are affected by the saturation effect as they were 
recorded for longer integration time, such that the averaged peak intensity 
levels were significantly lower. In order to mitigate the effect of saturation 
on the subsequent analysis, firstly additional contribution from saturation are
included in the error estimation of the flux scaling, that may arise from the 
underestimation of the average intensity of the profile. Secondly, in the 
spectral analysis of the time samples reported in the subsequent sections we 
consider only the conal regions that have much lower intensity levels and are 
free from saturation effect.

\subsection{Measuring Polarized Single Pulse Emission}

The Band-3 and Band-4 receiver systems of GMRT are equipped with dual linear 
polarization feeds that are converted into left and right hand circular 
polarization signals using a quadrature hybrid. The emission from PSR 
J0953+0755 was also recorded during these observations and served as a 
polarization calibrator. The auto and cross polarized signals of each time
sample were suitably calibrated to produce the four Stokes parameters (I, Q, U,
V) for each frequency channel \citep[see][for details]{2016ApJ...833...28M}. 
Finally, using the dispersion measure, DM = 26.7641~pc~cm$^{-3}$, and rotation 
measure, RM = -64.33 rad~m$^{-2}$, of PSR J0332+5434 the systematic changes in
the polarization behaviour across frequencies were corrected and subsequently 
averaged to produce the single pulse polarimetric time series corresponding to
the 6 subbands. The effect of cross-coupling in the antenna feeds have not been
addressed during this analysis and can lead to systematic errors up to 10\% of 
the measured Stokes parameters. The four Stokes parameters of the single pulses
in each subband were flux calibrated using the scaling factors from 
interferometric measurements (see discussion in the previous subsection). The 
two GMRT subarrays recorded independently and thereby introduced an arbitrary 
phase shift between the PPA traverse of Band-3 and Band-4 measurements that was
estimated by careful visual inspection. The two Stokes parameters Q and U of 
one of the bands were rotated appropriately with this phase shift to align the 
PPA distribution across the entire frequency range. An overlay of the
average stokes parameters for all the subbands after correcting for
the above effects is shown in the right panel of Fig~\ref{fig:fluxpuls}.

\section{High Linearly Polarized Time Samples and their Spectral Nature}\label{sec3}

\begin{figure}
\gridline{\fig{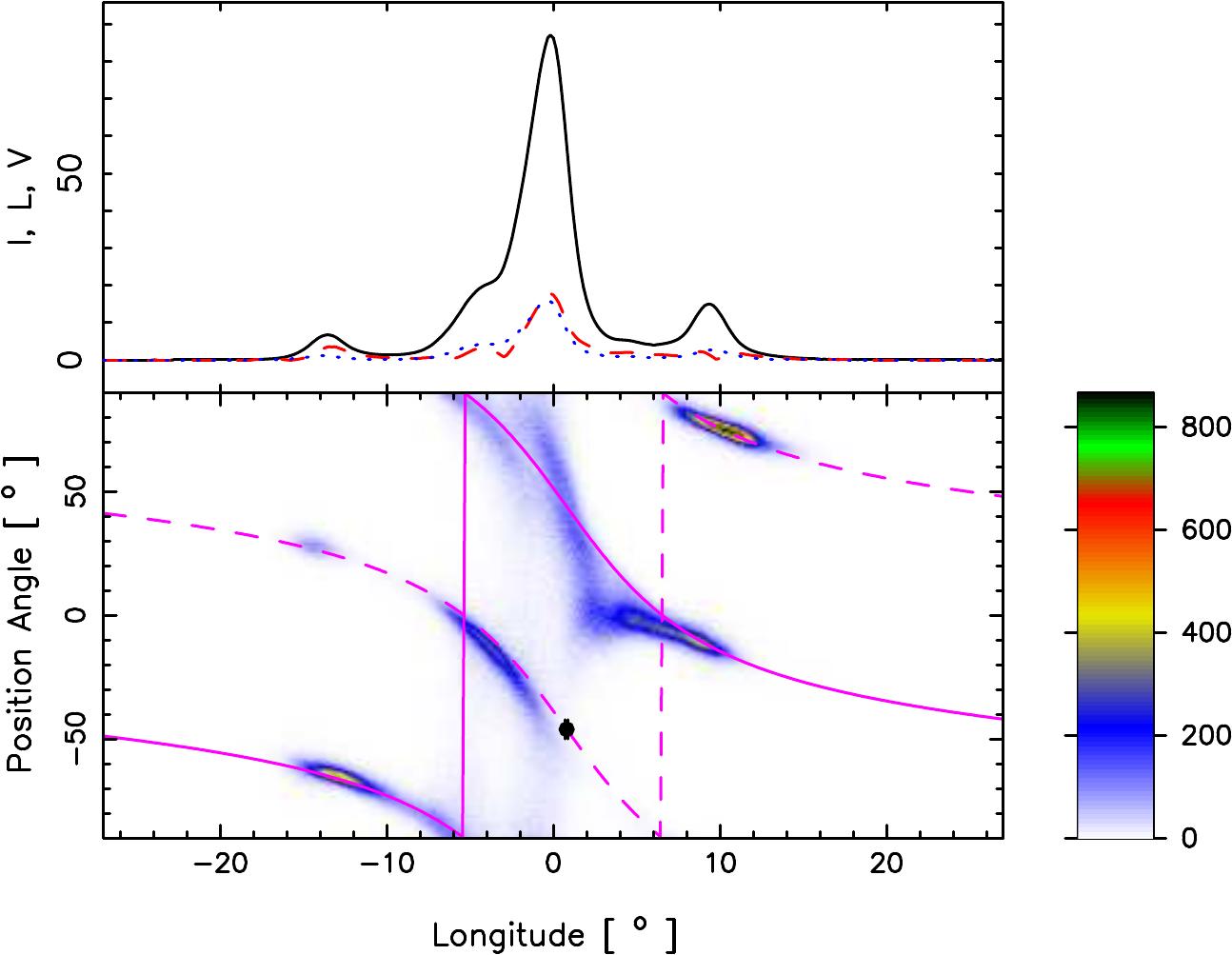}{0.32\textwidth}{}
          \fig{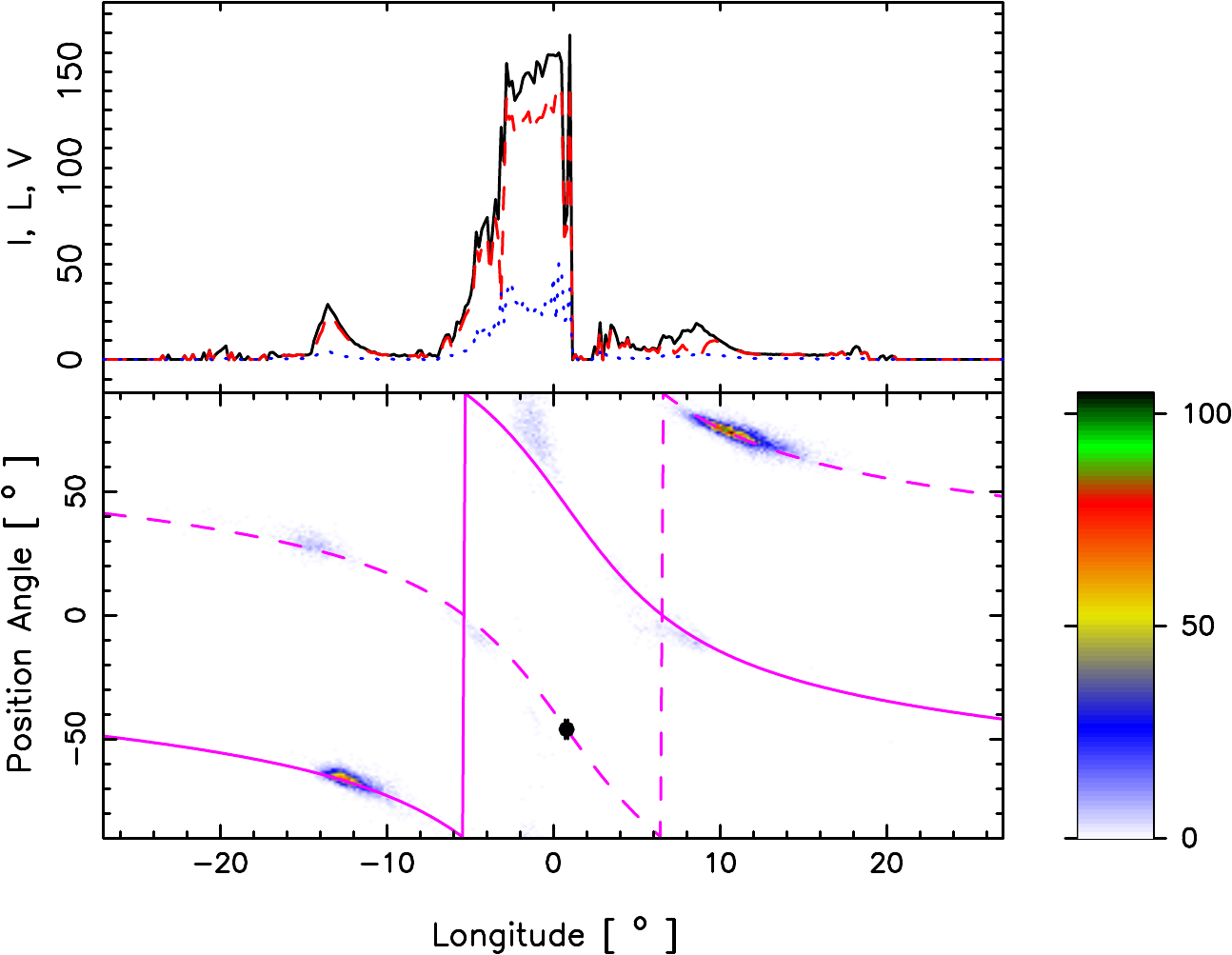}{0.32\textwidth}{}
          \fig{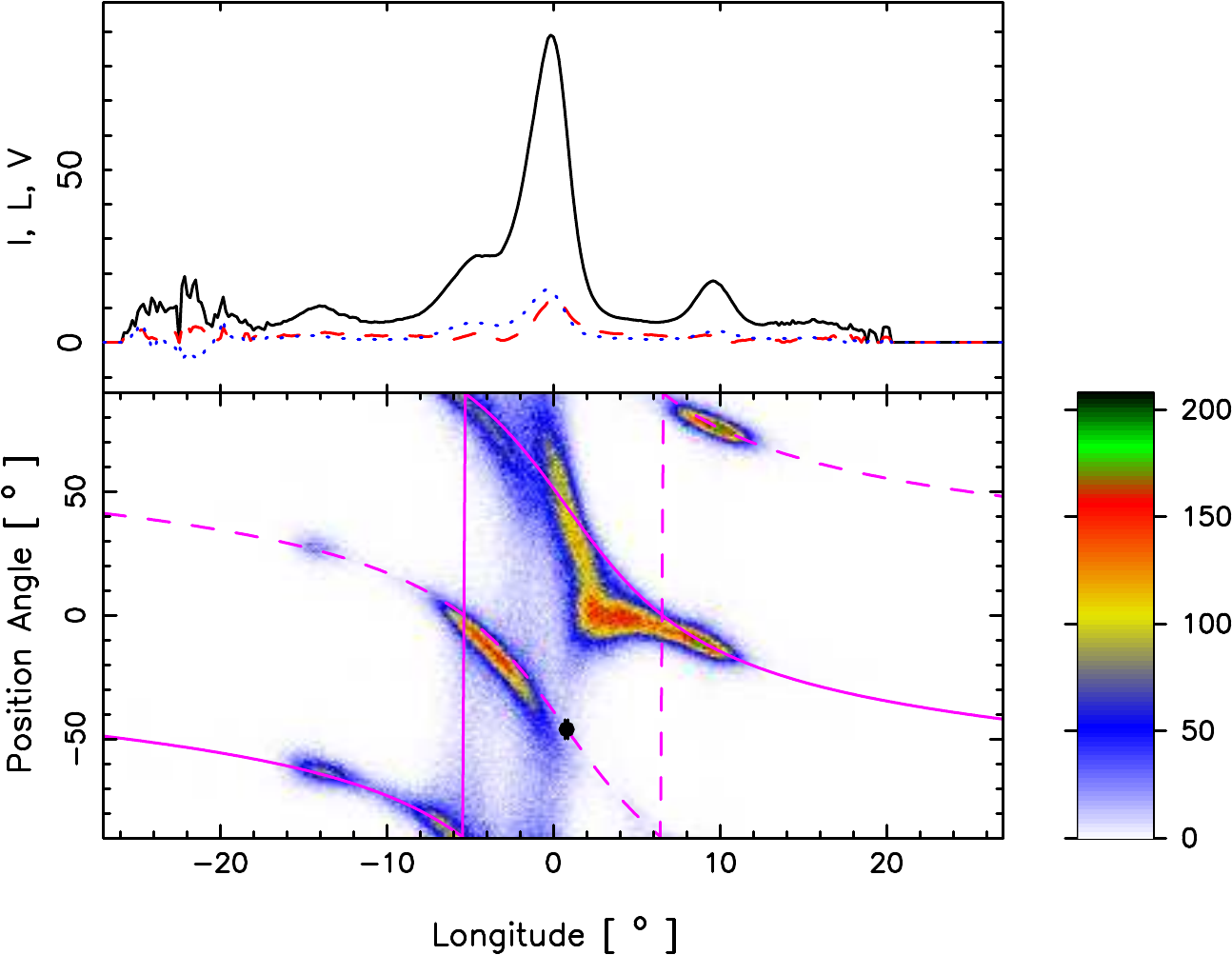}{0.32\textwidth}{}
	  }
\caption{The above plots show the polarization behaviour of PSR J0332+5434 
averaged over the entire 400 MHz frequency range, between 300 MHz and 750 MHz. 
The left panel comprises of all observed time samples, the middle panel 
corresponds to the high linearly polarized time samples with L/I $\geq$ 0.85, 
while the right panel includes the time samples with L/I below 50\%. The upper 
window in each panel shows the average profiles formed from the different time 
samples consisting of I (black), L (red) and V (blue). The bottom window shows 
the distribution of the PPAs of the time samples in colour scale along with the
RVM fits (magenta) for this pulsar obtained from the PPA distribution shown in 
Fig~\ref{fig4} (see discussion in section \ref{sec3}).
\label{fig3}}
\end{figure}

\begin{figure}
\gridline{\fig{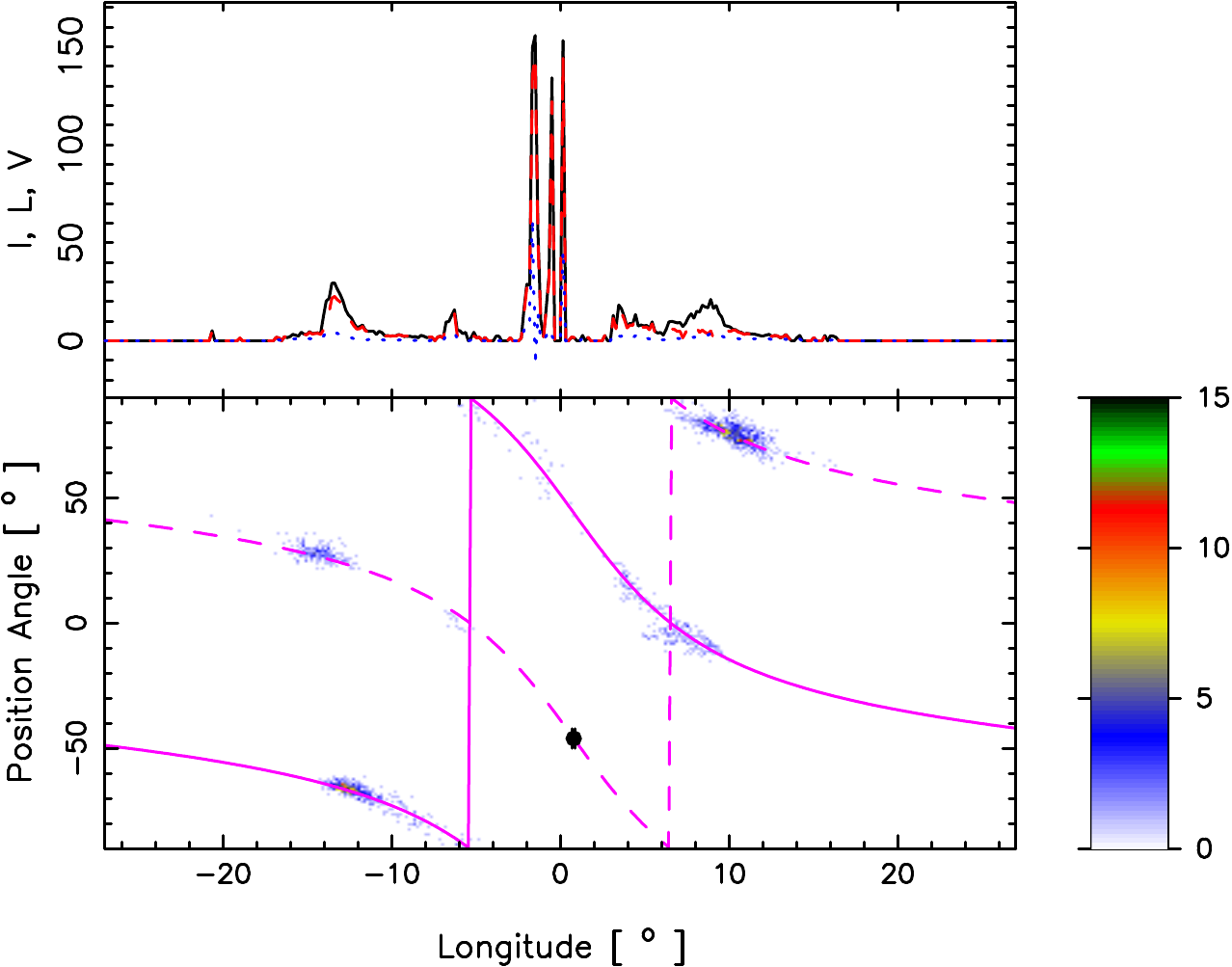}{0.44\textwidth}{}
          \fig{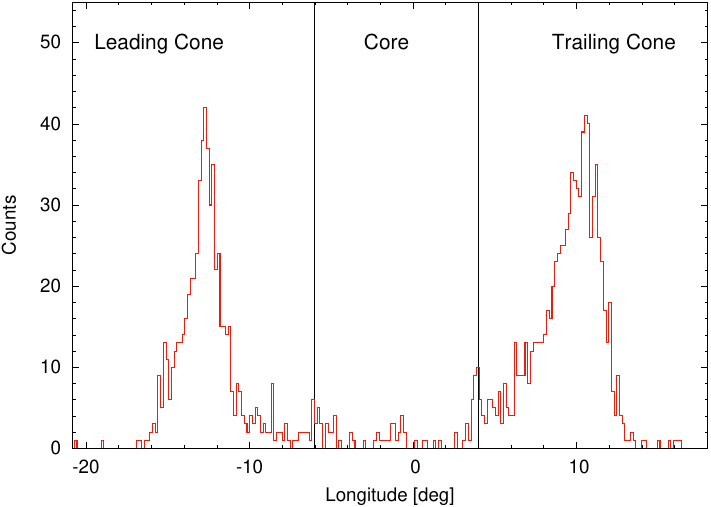}{0.48\textwidth}{}
	 }
\caption{The figure shows the polarization behaviour of PSR J0332+5434 averaged
over 400 MHz bandwidth between 300 MHz and 750 MHz. The figure uses time 
samples with L/I above 85\% in single pulses with weaker emission in the core 
region, i.e. less than one tenth of the maximum detection sensitivity. The left
panel is similar to Fig.~\ref{fig3} (see caption therein for details), while 
the right panel shows the number of time samples in each longitude bin used for
the left plot, with clear demarcation for the longitude ranges corresponding 
	to the leading cone, core and trailing cone shown in the figure.
\label{fig4}}
\end{figure}

Fig.~\ref{fig3}, left panel, shows the polarization behaviour of PSR J0332+5434
averaged for 400 MHz within the 300 MHz and 750 MHz frequency range.
The peaks of the 6 subbands were aligned before averaging (see Fig. 
\ref{fig:fluxpuls}, right panel) and included time samples with significant 
polarization detection, i.e. linear polarization level above five times the 
baseline noise rms of each single pulse. The two parallel RVM like PPA tracks 
can be readily seen in the lower window, although some PPA samples also spread 
out between the tracks and there is significant deviation from RVM in the core 
region, which has also been noted in \cite{2004A&A...421..681E}. 
\citet{2023MNRAS.521L..34M} showed that the PPA traverse of highly 
polarized time samples can uncover the underlying RVM nature in pulsars with 
complicated PPA distribution. The middle panel of Fig.~\ref{fig3} picks out the
high linearly polarized time samples, L/I exceeding 85\%, where the 
PPAs are mostly along the two orthogonal RVM 
tracks. The right panel of Fig.~\ref{fig3} comprises of the low polarization 
signals, L/I below 50\%, with the orthogonal RVM nature in the PPA 
distribution is discernible amidst the non-orthogonal behaviour. 
Initial results identifying the high linearly polarized emission from these 
observations over a limited frequency range, between 550 MHz and 750 MHz, have
been previously reported in \citet{2024Univ...10..248M}. These results are 
consistent with the general trend seen in the normal pulsar population 
\citep{2024MNRAS.530.4839J}. 

Although, the spread in the PPA distribution reduces in the high linearly 
polarized profile (middle panel, Fig.~\ref{fig3}), there still remains 
considerable deviation from the RVM nature below the core window within -6$^{\circ}$and 
$5^{\circ}$. 
\citet{2007MNRAS.379..932M} showed that the distortion of the PPA distribution 
in core region disappears in the single pulses where the core intensity is 
weaker. The high linearly polarized average profile is reconstructed by 
considering the single pulses with lower intensity core emission, i.e. less 
than 10\% of the maximum detection sensitivity recorded in the core region of 
the single pulses, and shown in Fig.~\ref{fig4}, left panel. Note that although 
our selection criteria was set to choose weak flux points in the core region, still 
about 10 time samples with high flux got selected. This is due to the fact that
the noise level in some single pulses were extrememly high and hence the 
the signal to noise for these high flux samples were low and fitted the selection criteria. 
We have retained these points since they do not affect the PPA distribution.
The overall distortion in
the PPA distribution is significantly reduced in the core region and they 
follow two parallel tracks. This updated distribution is used to estimate the 
RVM fits for PSR J0332+5434 at the observing frequency using the equation,
\begin{equation}
\Psi = \Psi_{\circ} +
	\tan^{-1}{\left(\frac{\sin{\alpha} \sin{(\phi-\phi_{\circ})}}
{\sin{(\alpha + \beta)} \cos{\alpha} - \sin{\alpha} \cos{(\alpha + \beta)}\cos{(\phi-\phi_{\circ})}}\right)}.
\label{eq1}
\end{equation}
Here $\Psi$ is the PPA from RVM, $\phi$ is the rotational phase, $\alpha$ is 
the angle between the rotation and magnetic axes, and $\beta$ is the angle 
between the rotation axis and the observer's line of sight during their closest
approach. $\Psi_{\circ}$ and $\phi_{\circ}$ defines the reference points for 
measurement offsets in $\Psi$ and $\phi$, respectively. 
We have used eq.(\ref{eq1}) to find appropriate fits of the emission 
geometry from the measured PPA using $\chi^2$ minimization techniques and 
obtain the minimum reduced $\chi^2 \sim 6.5$ with corresponding values of the 
parameters as $\alpha = 46.8^{\circ}$, $\beta = -4.5^{\circ}$ and $\phi_{\circ}
= 0.8^{\circ}\pm0.4^{\circ}$. However, it has been well established that the 
geometric angles $\alpha$ and $\beta$ obtained from RVM fitting are highly 
correlated and should be treated with caution (see e.g. 
\citealt{2001ApJ...553..341E}). The orthogonal RVM fits to the PPA distribution
are shown in as Fig~\ref{fig3} as well as the left panel of Fig.~\ref{fig4} 
(magenta curves). 
The solid magenta curve specifies the primary polarization mode (PPM) 
corresponding to $\psi_{\circ} = 44^{\circ}\pm5^{\circ}$, while the orthogonal 
dashed curve represents the secondary polarization mode (SPM) with 
$\psi_{\circ} = -46^{\circ}\pm5^{\circ}$.
The right panel in Fig.~\ref{fig4} presents the number of time samples in each 
longitude range where L/I exceeds 85\%, with the added constraint of single 
pulses having lower intensity core emission.

\subsection{Selecting Single Pulse Emission Features for Spectral Studies}
In the observing frequency range, between 300 MHz and 750 MHz, the single pulse
emission have similar shapes indicating a broadband nature, which is consistent
with the results of K03. There are several examples of high linearly polarized 
signals in the single pulses seen across all 6 subbands. But there is large 
variability in single pulse structure often in the form of sporadic 
illumination at different regions of the emission window. Both wider features 
spanning several degrees of longitude and narrow spiky emission about a degree 
in width is seen in different pulses. The spiky features are often highly 
polarized and appears at the same longitude range at all frequencies, and 
due to their relative ease of identification we particularly focus on their 
spectral nature in the subsequent analysis. 

\begin{figure}
\epsscale{1.05}
\plottwo{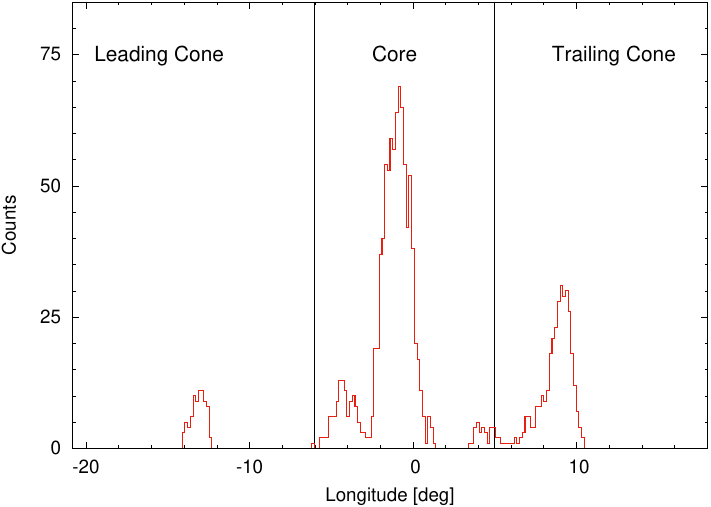}{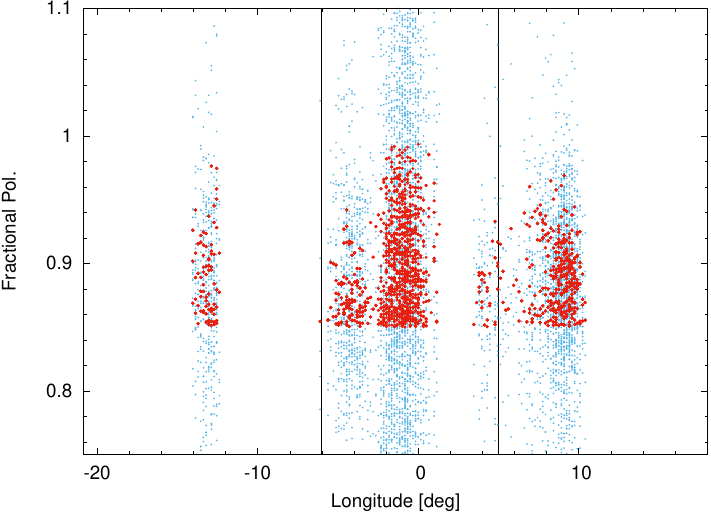}
\caption{The selection criteria for the band averaged time samples with high 
levels of linear polarization are, L/I exceeds 85\% and they have significant 
detection sensitivity, exceeding 15 times the noise rms levels in the single 
pulse baselines. The left panel in the figure shows the number of points in 
each longitude range that satisfy the selection criteria. The right panel shows
the linear polarization fractions of these time samples across the profile 
window, with the red points representing the average over the entire frequency 
range, while the blue points correspond to the 6 subbands showing a larger 
spread than the average. The longitude windows of the leading cone, core and 
trailing cone are also shown in the figure.
\label{fig5}}
\end{figure}

The high linearly polarized time samples in the single pulses have been 
identified in Fig~\ref{fig3}, middle panel, from the 400 MHz band between 
300 to 750 MHz, and constitute L/I $\geq$ 85\% and significant linear 
polarization detection sensitivity above 5 times the noise rms levels of the 
single pulse baselines. However, in the individual subbands the linear 
polarization levels of these time samples show a larger spread, often due to 
higher noise levels in each subband and consequently lower detection 
sensitivity. In order to minimise the spread the selection criterion is 
tightened to include time samples with detection sensitivity above 15 times the
baseline noise levels and at least five times above the noise levels for each 
individual subband. Fig.~\ref{fig5}, left panel, shows the number of time 
samples at each longitude bin that satisfy this section criterion. We have 
limited the spectral studies to the conal region of the pulsar profile, 
excluding the core between longitude range of -6$\degr$~and 5$\degr$ (see
discussion in the previous section about saturation in the core). In some 
instances the L/I exceeds 100\% which is unphysical, but in realistic 
measurements this can be attributed to cumulative errors like statistical 
fluctuation in the linear polarization, improper removal of polarization 
baseline levels, high baseline noise, cross-coupling between different 
polarizations, etc. Attributing an additional 10\% error due to these effects
the spectral studies are restricted to time samples in 6 subbands with L/I 
between 75\% and 110\%. Fig.~\ref{fig5}, right panel, shows the distribution of
the L/I for the selected time samples across the profile window.

\begin{figure}
\epsscale{0.8}
\plotone{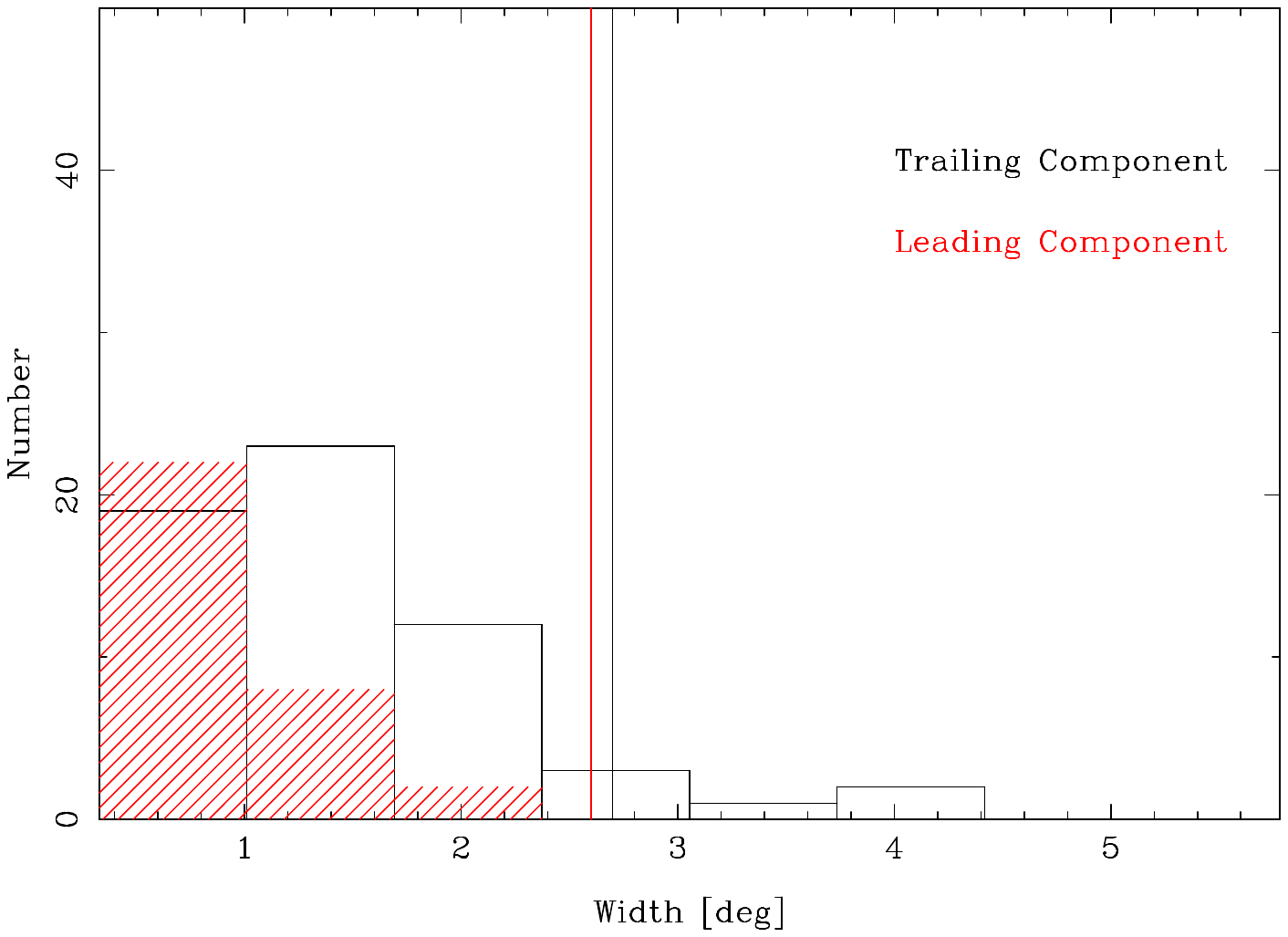}
\caption{The figure shows the distribution of the widths of the highly 
polarized emission features in PSR J0332+5434 measured over 400 MHz band,
between 300 MHz and 750 MHz, and obtained using our selection criteria (see 
text for details). The red distribution corresponds to features seen in the 
leading conal window while the black histogram represents the features in the 
trailing conal side. The widths were estimated at 50\% level of the peak 
intensity. The corresponding widths for the leading and trailing cone in the 
average profile is shown as vertical red and black lines, respectively.
\label{fig6}}
\end{figure}

The above selection criteria yielded a total of 93 high linearly polarized
time samples, with about 34\% samples belonging to the leading cone and 66\% to
the trailing conal component. The widths of the emission features associated 
with the high linearly polarized time samples, obtained from the full band, 
have been estimated at 50\% level (W$_{50}$) of their peak intensity, and the 
distribution of the widths corresponding to the leading and trailing cones is 
shown in Fig.~\ref{fig6}. More than 70\% of these are narrow spiky
features with widths less than 1.5$\degr$~in longitude (see examples in Fig.
\ref{fig8}), compared to around 3$\degr$ for the leading and trailing conal
components in the average profile. In the leading conal side around 90\% of the
highly polarized features belong to the narrow spiky category, while there is
more even distribution in the trailing cone, with around 60\% narrow features
and the remaining 40\% wider features with widths in excess of 1.5$\degr$. The
spiky emission is localised within a narrow longitude range across all 6
subbands with their peaks being mostly coincident.

The narrow localized nature of the 
spiky emission bears resemblance to the quasi-periodic emission structures found
in single pulses of certain pulsars, known as microstructures 
\citep{1983ApJ...272..687B,1987AZh....64.1013P,1994A&A...286..807S}. Several
studies have found strong correlation between microstructure widths ($w_{\mu}$)
and the pulsar period with a dependence, $w_{\mu} \approx 0.5 \times 10^{-3} P$
\citep{1975ApJ...195..513T,1979AuJPh..32....9C,2002MNRAS.334..523K,
2015ApJ...806..236M}. Using this relation the expected microstructure width in 
PSR J0332+5434, with $P = 0.714$ seconds, is $w_{\mu} \sim 357$ $\mu$s. This is
comparable to the observing time resolution, 327 $\mu$s, and hence 
microstructures cannot be detected from these measurements. Additionally, the 
spiky highly linearly polarized emission is much wider than the expected 
microstructures and appear to be unrelated to them.

\subsection{Spectral Nature of Highly Polarized Emission}

\begin{figure}
\gridline{\fig{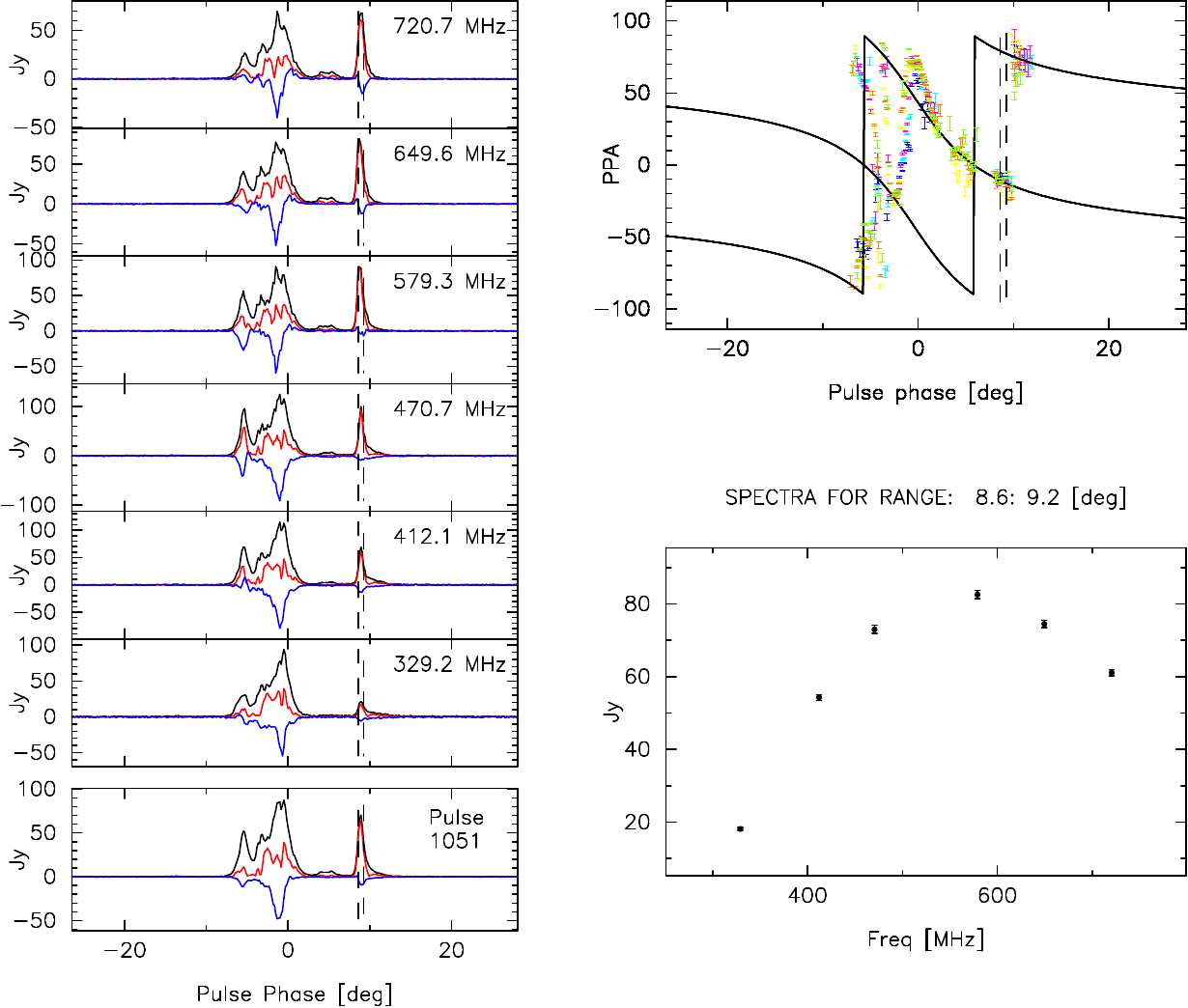}{0.44\textwidth}{(a) Pulse number : 1051}
          \fig{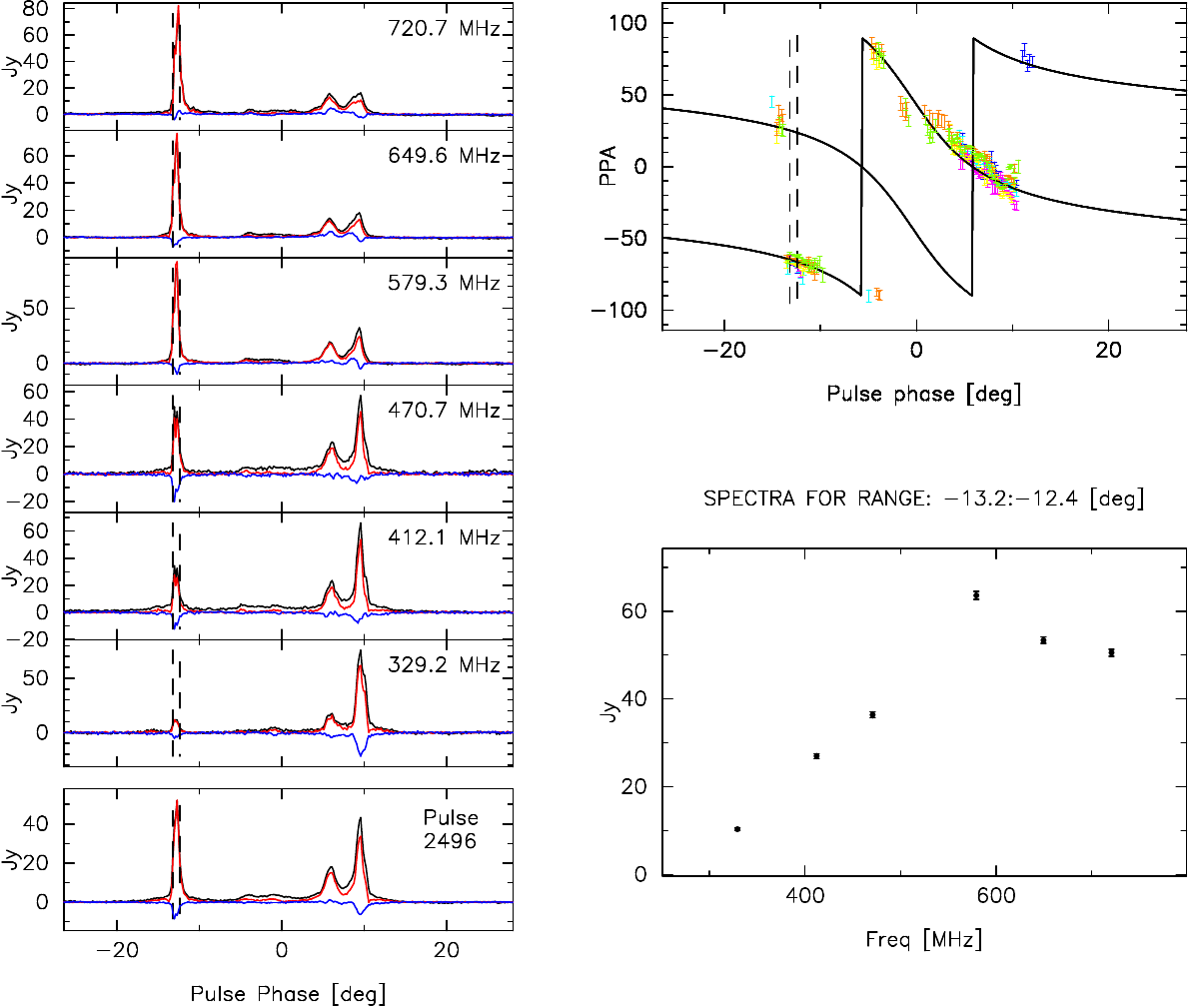}{0.44\textwidth}{(b) Pulse number : 2496}
	  }
\gridline{\fig{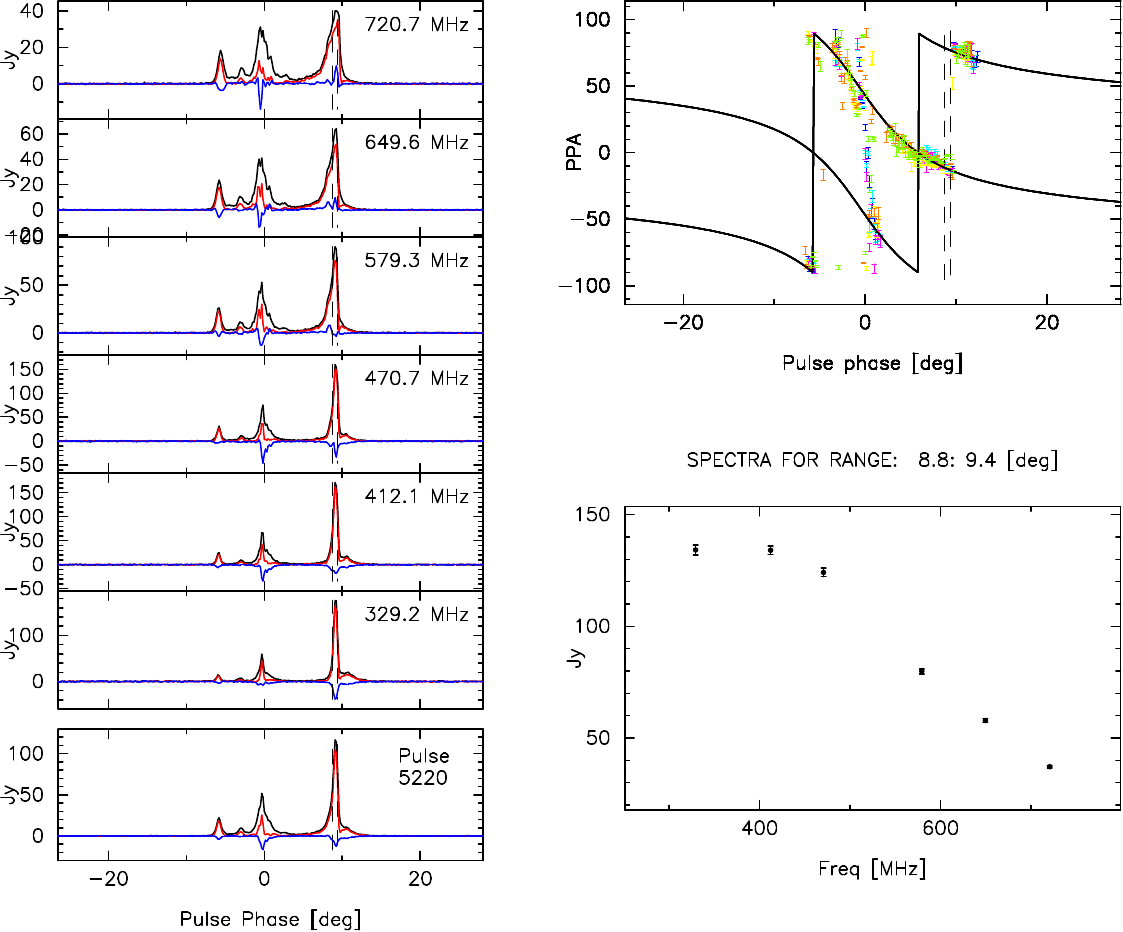}{0.44\textwidth}{(c) Pulse number : 5220}
          \fig{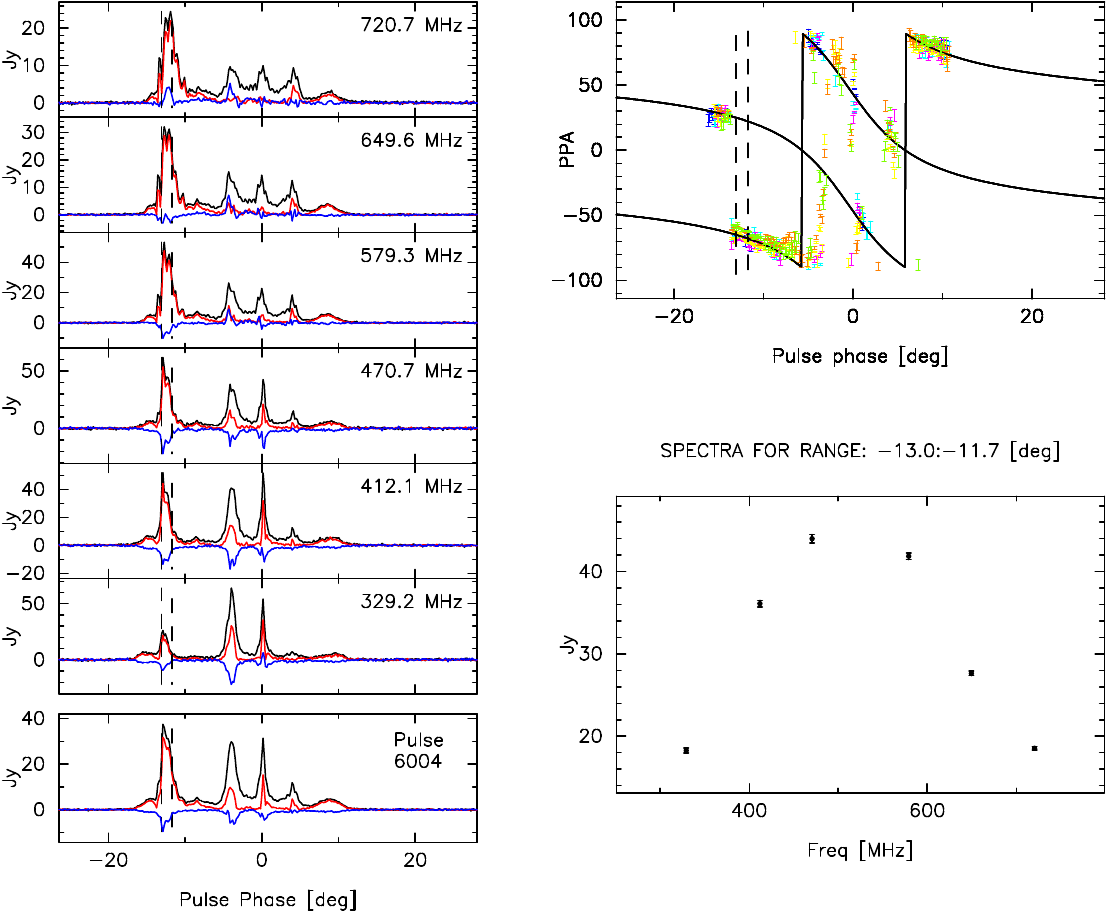}{0.44\textwidth}{(d) Pulse number : 6004}
	  }
\caption{The figure shows four single pulses of PSR J0332+5434 containing high 
linearly polarized spiky emission, exemplifying their polarization behaviour 
as well as spectral nature. Each panel comprises of three separate windows, 
where the left window shows the single pulse I, L and V at the 6 subbands 
between 300 MHz and 750 MHz as well as the average behaviour across the entire 
band at the bottom. The dashed lines in the figures identify the spiky emission
from the full band, specified by $W_{50}$ of the feature. The top window on the 
right shows the PPA distribution across the single pulse for all 6 frequencies 
(see caption in Fig.\ref{fig:fluxpuls} for the frequency colour code). The PPAs
of the spiky emission belong to the PPM in all cases. The bottom window on 
the right shows the spectral nature of the spiky polarized emission, where the 
average intensities at the 6 frequencies are estimated within the dashed 
window. (a) Pulse number 1051 where the spiky emission is seen in the trailing 
cone and the spectrum turns over with peak around 600 MHz. (b) Pulse number 
2496 with the spiky emission in the leading cone and the spectrum reverses to 
positive spectral index around 600 MHz. (c) Pulse number 5220 with the spiky 
emission in the leading cone and the spectrum flattens at lower frequencies 
below 400 MHz. (d) Pulse number 6004 where the leading cone contains the spiky 
emission that has a inverted parabolic spectrum peaked around 500 MHz.
\label{fig8}}
\end{figure}

\begin{figure}
\gridline{\fig{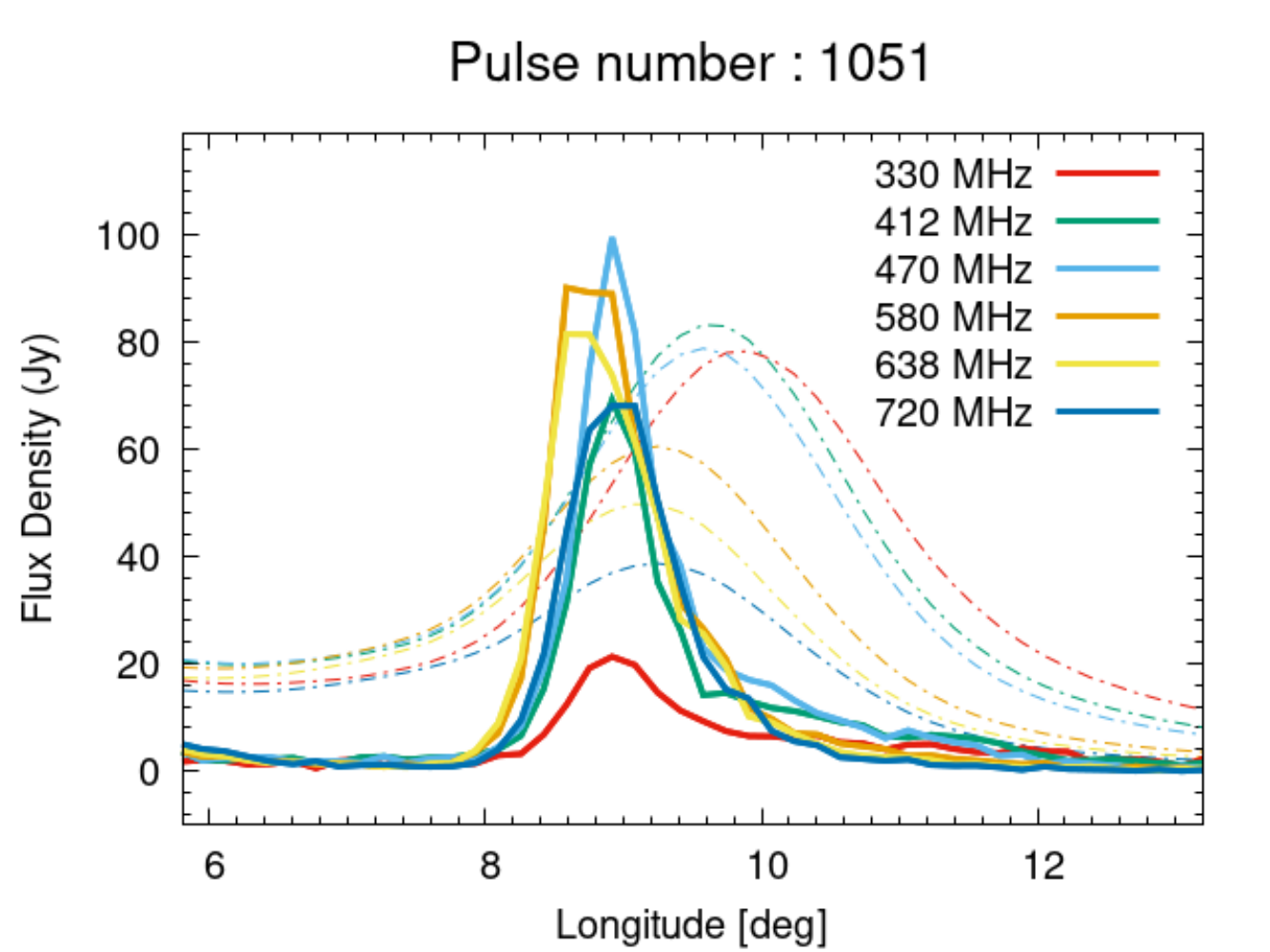}{0.45\textwidth}{}
          \fig{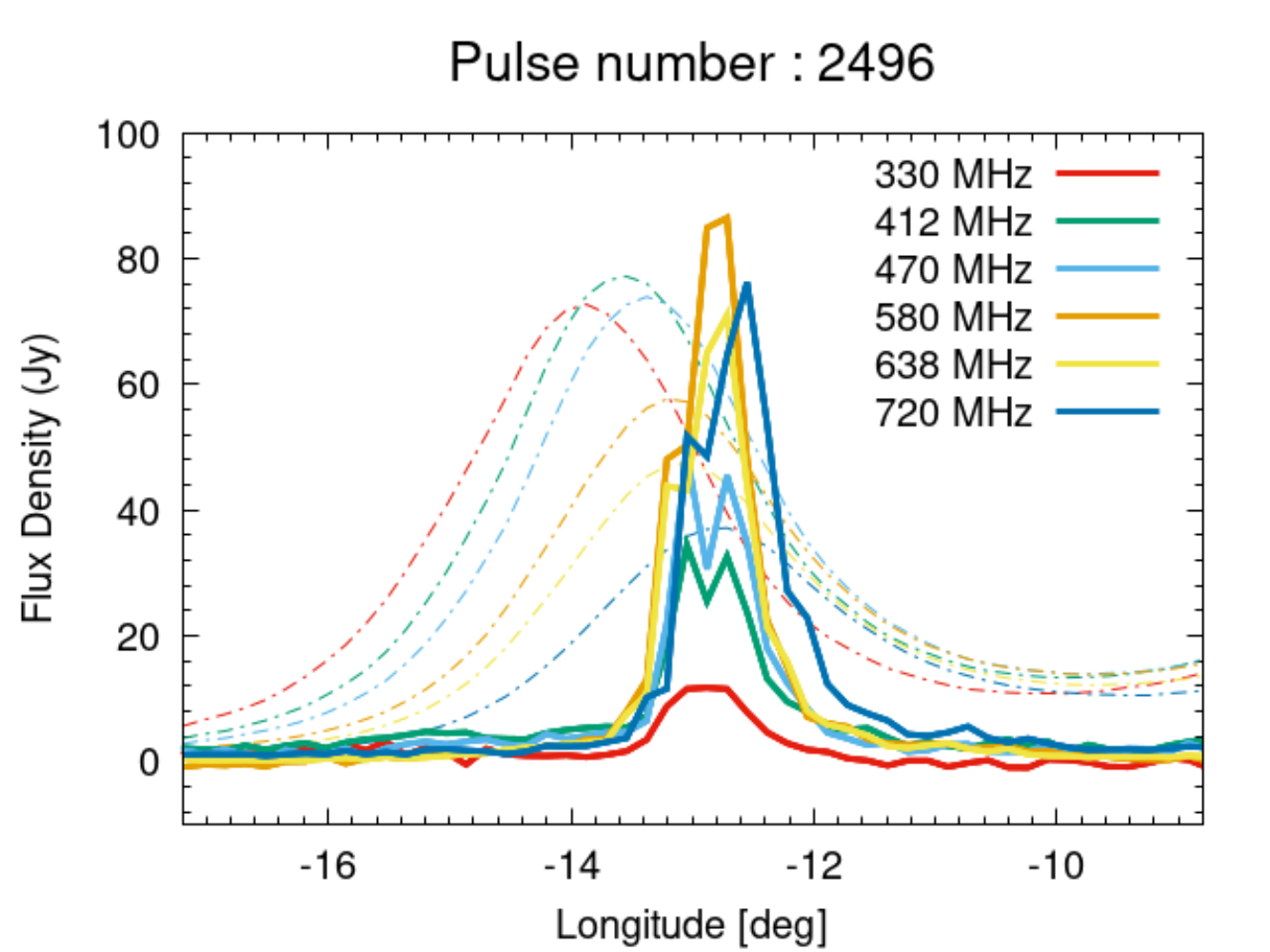}{0.45\textwidth}{}
          }
\gridline{\fig{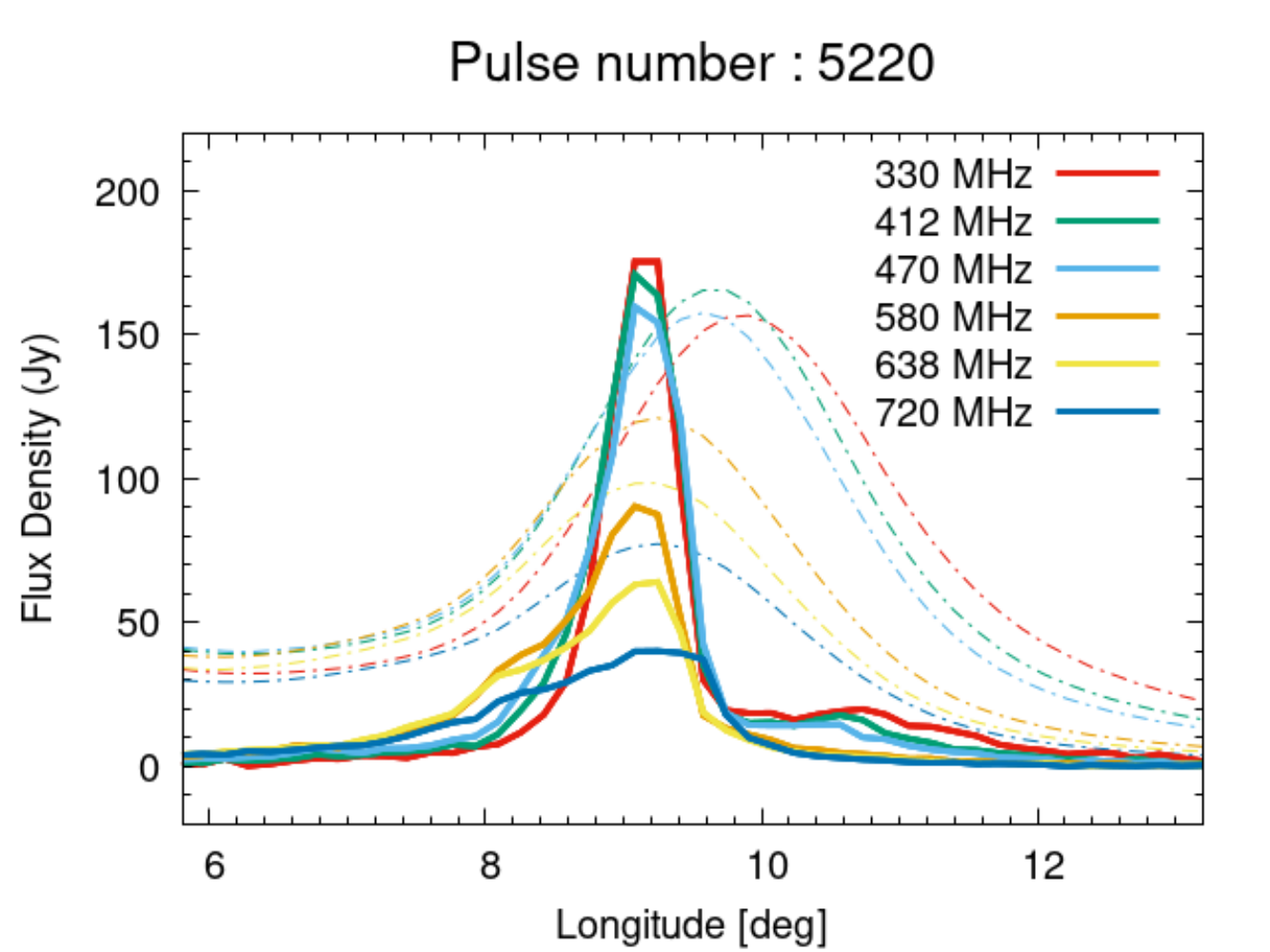}{0.45\textwidth}{}
          \fig{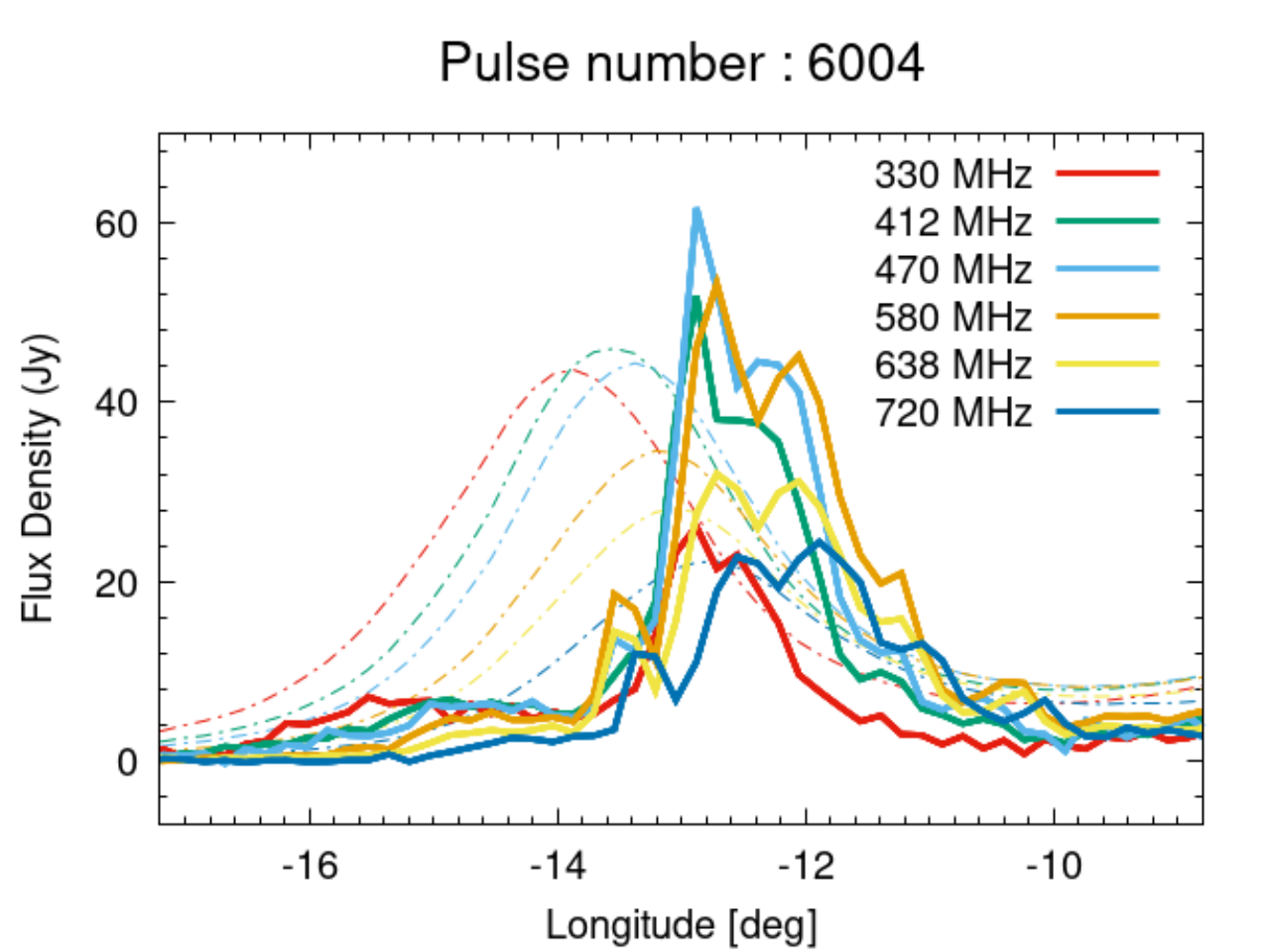}{0.45\textwidth}{}
          }
\caption{The figure shows an expanded view of the highly polarized spiky 
features in the four pulses reported in Fig.~\ref{fig8}, at frequencies 330 MHz
(red), 412 MHz (green), 470 MHz (blue), 580 MHz (orange), 638 MHz (yellow) and 
720 MHz (indigo). The dot-dashed lines in the background of each panel show 
the relevant conal components of the average profiles at these frequencies, 
trailing cone in the left panels and leading cone in the right panels, with 
the same frequency colour scheme, and intensities scaled appropriately to be 
visible within the window. The average profile as well as the single pulses in 
each frequency have been aligned such that the profile peak is at zero 
longitude. The effect of RFM is readily seen in the cones, i.e. the emission 
shifts outward from the center with decreasing frequency. But this phenomenon 
is absent in the spiky emission, where the emission windows at different 
frequencies show little shift usually within one or two longitude bins.
\label{fig9}}
\end{figure}

The spectra of highly polarized emission features are estimated using $W_{50}$ 
window from the full band to find the average flux density at each frequency. 
Fig.~\ref{fig8} shows the emission behaviour of the spiky highly polarized 
features in four single pulses, pulse numbers 1051, 2496, 5220 and 6004 from 
the start of the observing session. The left window in all four panels, show 
the intensity in the single pulses corresponding to I, L and V at all 6 
subbands as well as the full 400 MHz band, where the $W_{50}$ window has been 
identified with dashed lines. The spiky emission is seen within the trailing 
cone in pulses 1051 and 5220, but in the leading conal side in pulse 2496 and 
6004. The top window on the right side shows the PPA distribution at all 6 
frequencies while the bottom window represents the estimated spectrum obtained 
from the average flux estimates. 

The spiky features in all cases have a broadband character and appear almost in 
the same longitude range at all frequencies. This is highlighted in 
Fig.~\ref{fig9} showing the expanded view of spiky emission features across the
frequency range, overlaid on the relevant average conal component at these 
frequencies. The features are narrower than the cones and are largely localised
within a narrow longitude range. In contrast both the leading and trailing 
conal components clearly show the effect of radius to frequency mapping (RFM), 
according to which the profile width progressively increases with decreasing 
frequency. The average emission at different frequencies are expected to 
originate from different heights in the pulsar magnetosphere, with higher 
frequencies arising closer to the stellar surface. However, the lack of RFM in 
the spiky emission suggests that they are localised within the magnetosphere, 
with the same source emitting over the frequency range of detection.

The spectra of these features have a variety of shapes, the most noticeable
being the inverted spectra with a turnover frequency (see spectra of
pulses 1051 and 6004~in Fig.~\ref{fig8}) and the spectra falling on either side
of the turnover frequency resembling a inverted parabolic shape and further 
the spectra differs greatly from the average spectra of the cones (see right 
panel of Fig.~\ref{fig:avgcompspect}). The turnover frequencies in these 
inverted parabolic spectra are not fixed but shift within the entire band, e.g.
the turnover frequency is around 600 MHz in pulse number 1051 but shifts to 500
MHz in pulse 6004. We found 42 spiky features with inverted parabolic 
spectral shape, constituting around 60\% of the total sample. The spectra show 
flattening at the lower frequencies in several cases, while at other instances 
the spiky feature merge with surrounding emission and the resultant spectra in 
the $W_{50}$ window has irregular shape. The PPAs in all spiky features follow 
the RVM and are associated exclusively with the PPM.

\begin{figure}
\gridline{\fig{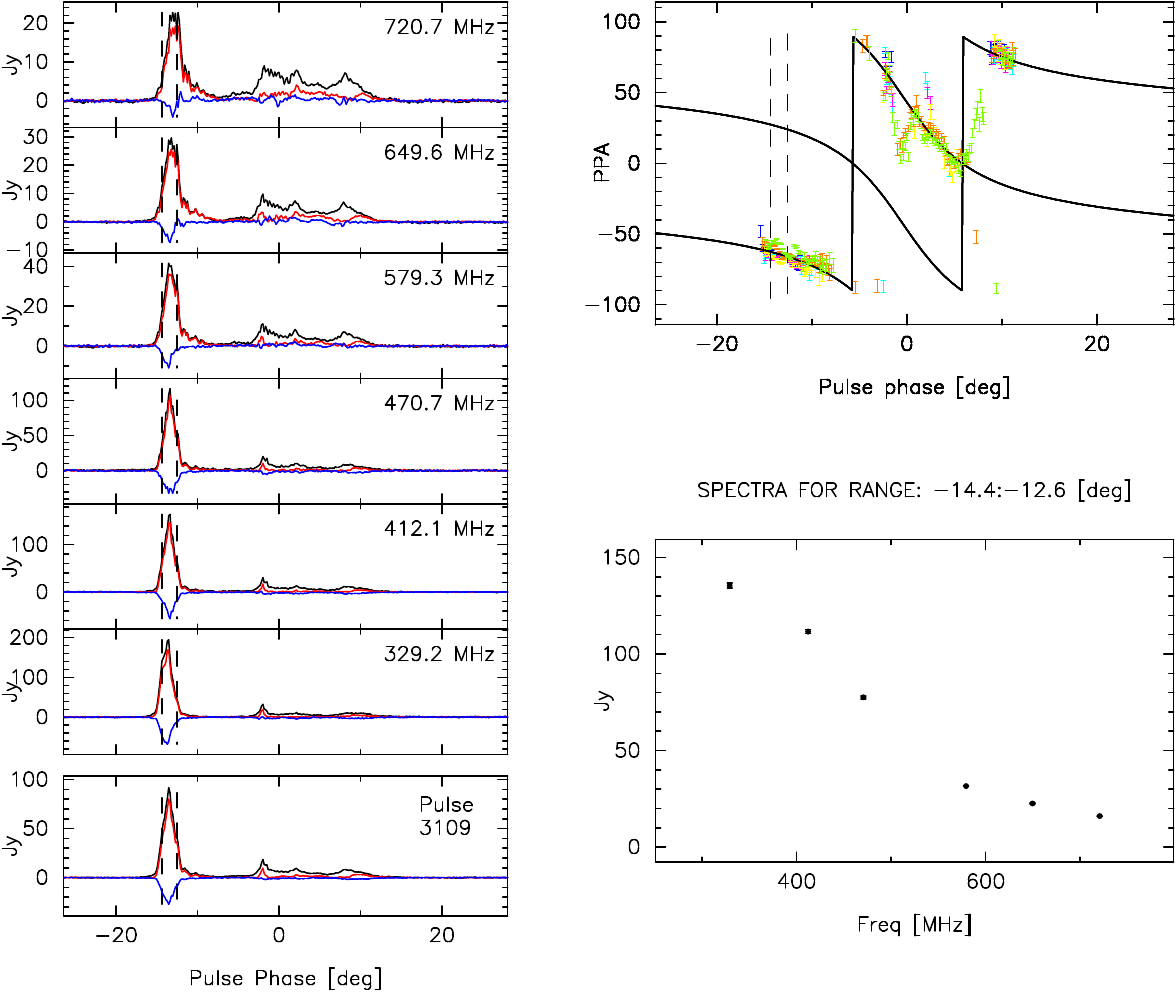}{0.45\textwidth}{(a) Pulse number : 3109}
          \fig{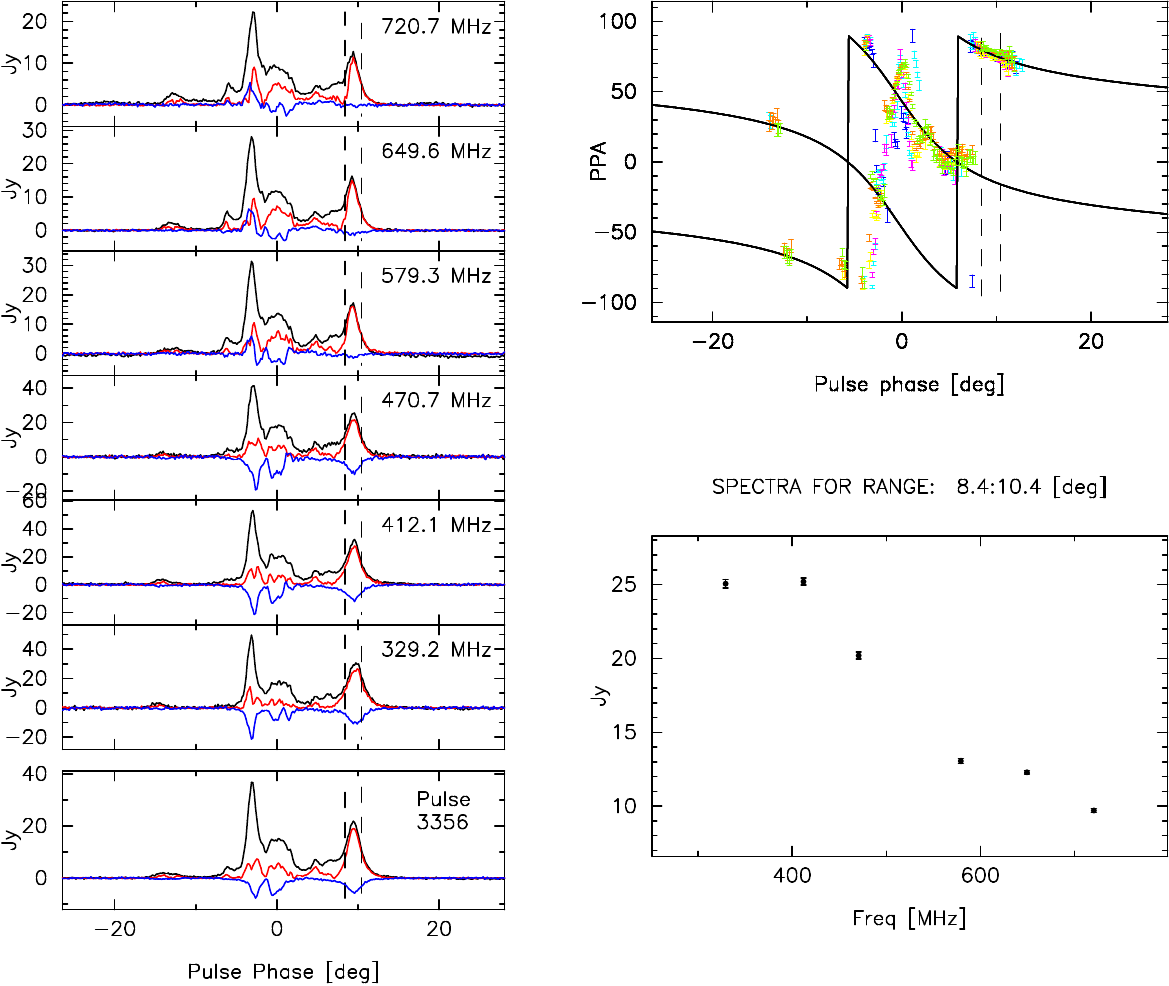}{0.45\textwidth}{(b) Pulse number : 3356}
          }
\caption{The above plot shows two examples of single pulses from PSR J0332+5434
containing high linearly polarized broad emission features. The details of the
individual panels are identical to Fig.~\ref{fig10} (see caption there for
description). However, unlike the spiky emission the PPAs of the wide features
are seen along both the PPM and the SPM. (a) Pulse number 3109 where the broad 
highly polarized emission is seen in the leading cone and the spectrum is 
mostly inverted power law with flattening above 600 MHz. (b) Pulse number 3356 
where the broad emission is seen in the trailing cone and the spectrum has 
irregular shape.
\label{fig10}}
\end{figure}

\begin{figure}
\gridline{\fig{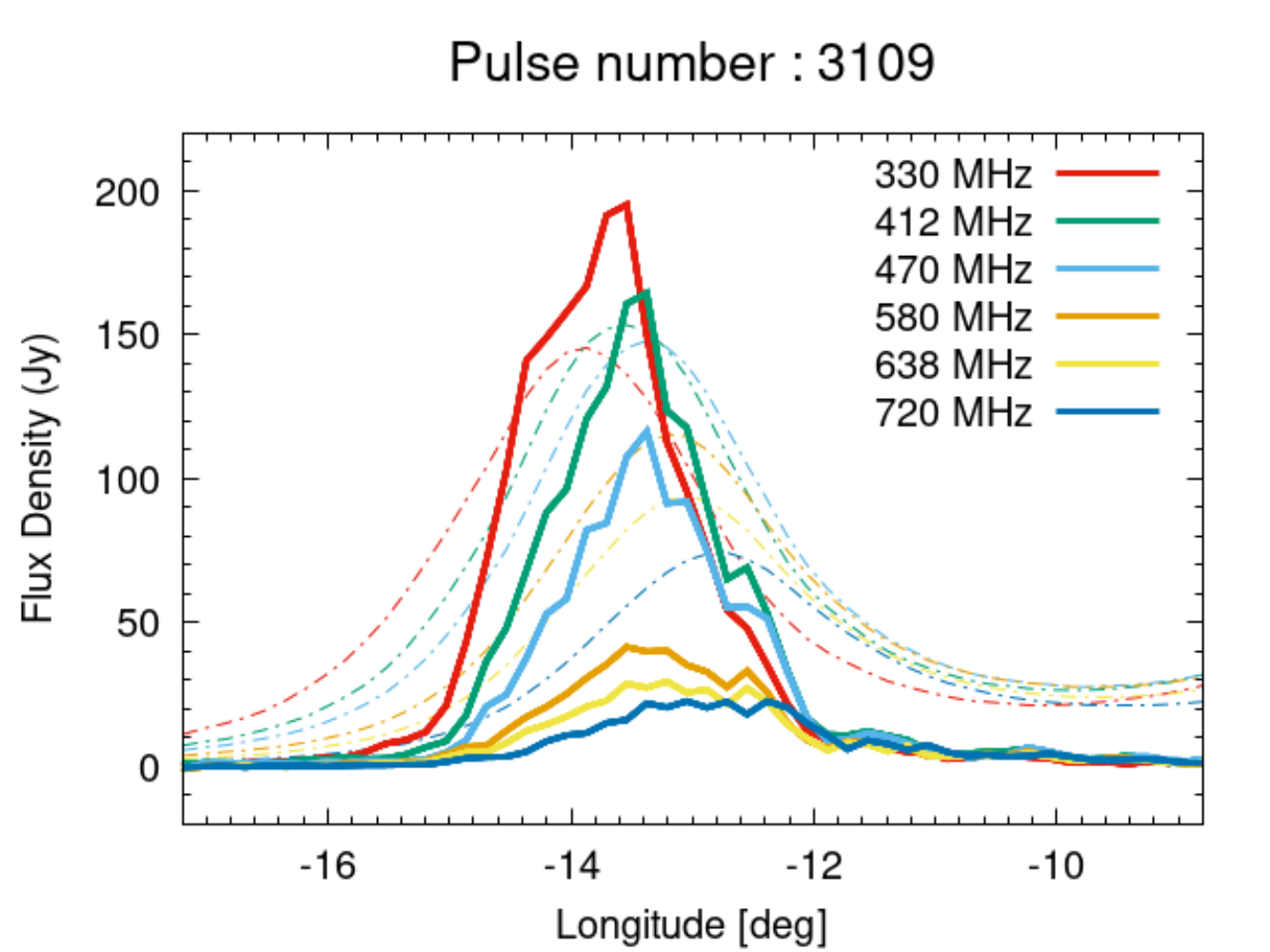}{0.45\textwidth}{}
          \fig{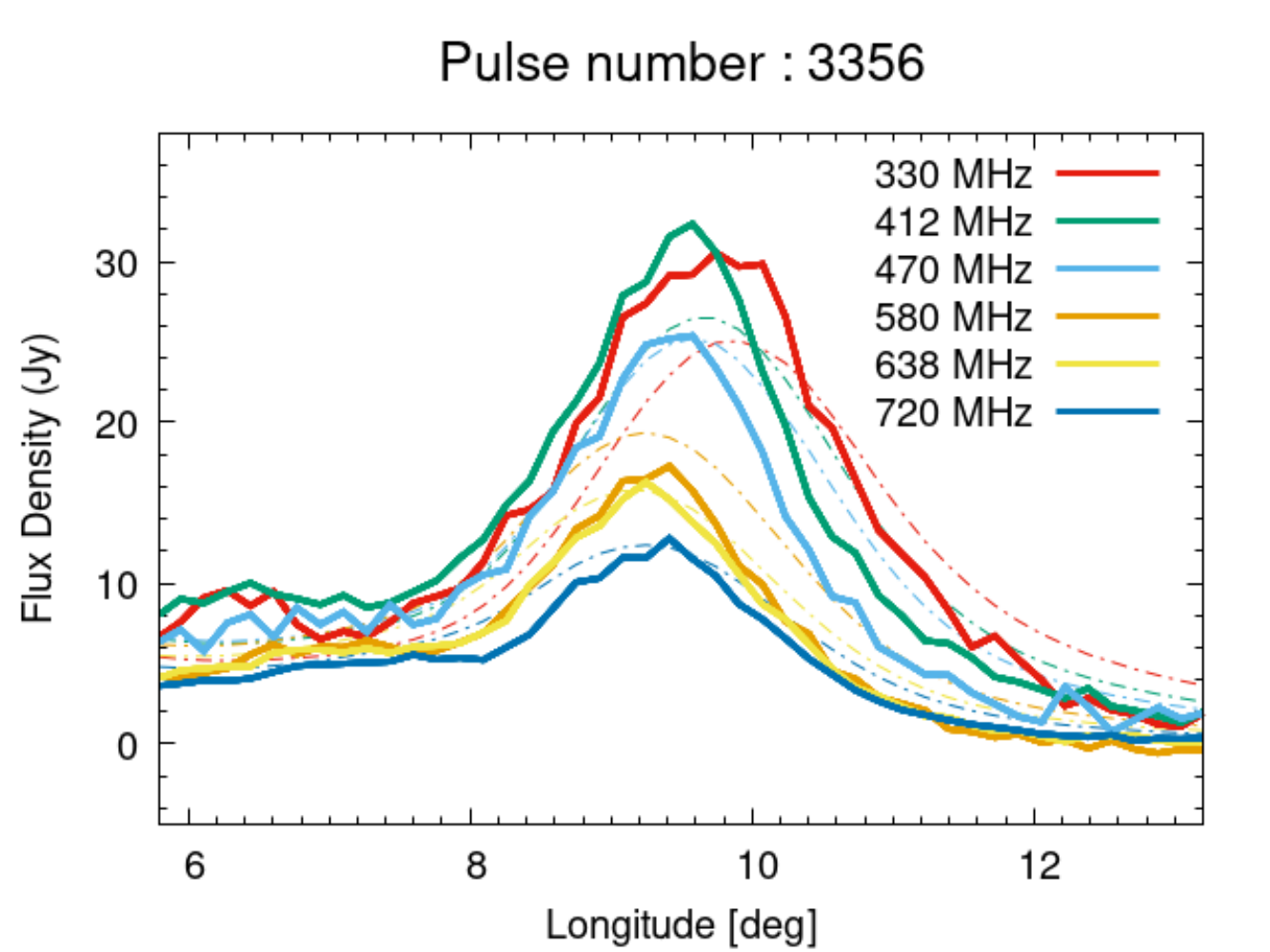}{0.45\textwidth}{}
          }
\caption{The figure shows an expanded view of the highly polarized broad 
features in the two pulses reported in Fig.~\ref{fig10}. The figure is 
identical to Fig.~\ref{fig9} (see caption for description). The broad features
mostly match the radius to frequency mapping of the conal components.
\label{fig11}}
\end{figure}

In addition to the spiky features there are several examples of broader 
features with W$_{50} > 1.5\degr$, that exhibit high levels of L/I. Examples of
these features are seen in pulse number 3109 at the location of the leading 
cone and in trailing conal side of pulse number 3356, as shown in Fig.
\ref{fig10}. The wider features are clearly different from the spiky emission 
in their physical properties. There are a few examples of
inverted parabolic spectral shape even in these wider features, but they constitute less
than 20\% of the total sample. In most instances the spectra have irregular
shape, but the flux generally decreases with increasing frequency. A closer
look into the wider features shown in Fig.~\ref{fig11} reveal that they largely
follow the tendency of RFM, consistent with the average conal behaviour.
Although the PPAs in these features follow the RVM, they can be associated with
both the PPM and the SPM.

\section{Spectral nature of high linearly polarized emission : Qualitative Concepts}\label{sec4}

The polarization behaviour is expected to be a direct consequence of the 
underlying radio emission mechanism in pulsars. But the observed spectra is
likely to be different from the one obtained from the intrinsic emission 
mechanism as they get reshaped through a number of subsequent processes, like 
averaging of emission from multiple sources with shifted spectra, propagation 
in the pulsar magnetosphere as well as the intervening medium, the observers' 
line of sight geometry, etc. In the remainder of this section we will carry out
a qualitative discussion about the origin of polarization from the emission 
mechanism as well as the effect of different phenomena on the observed spectra 
from pulsars, starting from the role of the emission mechanism. A more detailed
quantitative study to model the observed average spectra as well as from the 
different single pulse polarized features will be deferred to a future 
dedicated work \citep[see][for details]{2022ApJ...927..208B}.

The first and most important step towards understanding the radio emission 
mechanism is determining the location of the emission region in the pulsar
magnetosphere. A widely used technique for estimating the radio emission height 
utilizes the aberration and retardation (A/R) effect arising due to rotation of
the pulsar, which causes a shift ($\Delta \phi$) of the profile window with 
respect to the magnetic axis. The emission height, $h$, is related to this 
shift as $\Delta \phi = 1440 h/P c$, where $\Delta \phi$ is in degrees, 
$c$ is the velocity of light and $P$ is the pulsar period 
\citep{1991ApJ...370..643B}. The longitude difference between the profile 
center and the SG point ($\phi_{\circ}$) of the PPA traverse gives measure of 
$\Delta \phi$ from observations. As shown earlier, the PPAs of PSR 
J0332+5434 can be used to obtain RVM fits with estimates of $\phi_{\circ} = 
0.8\degr\pm0.4\degr$. The outer edges of the average profile from the full
band, estimated at 10\% level of the conal components, are located at $\phi_l =
-16.3\degr\pm 0.4$ in the leading side and $\phi_t = 12.7\degr \pm 0.4$ in the 
trailing side. The profile center is located at -1.8$\degr$ and using $P = 
0.715$ seconds, gives estimate of the average profile height to be $h\sim386 
\pm 85$ km, which is well below 10\% of the light cylinder radius ($Pc/2\pi 
\sim 35000$ km). The PPM and SPM fit orthogonal RVM models with identical 
$\phi_{\circ}$, suggesting similar emission heights for both modes.

The pulsar magnetosphere is filled with a dense electron-positron pair plasma 
streaming outwards in a relativistic non-stationary flow along the open field 
lines. The estimated emission height suggests the coherent radio emission to be 
excited in the strongly magnetized limit, and if we consider identical 
distribution functions of electrons and positrons, then the dispersion 
equations describing this system yield two orthogonal linearly polarized modes,
the ordinary (O-mode) and extraordinary (X-mode) mode. The polarization vector 
of O-mode lies in the plane of propagation vector, $\vec{k}$, and the ambient 
magnetic field, $\vec{B}$, with a component along $\vec{B}$ as well as 
$\vec{k}$. The X-mode, has transverse character where the polarization vector 
is directed perpendicular to the plane containing $\vec{k}$ and $\vec{B}$, and 
is also known as $t$-mode\footnote{For definition of plasma modes we use the convention 
by \cite{2003PhRvE..67b6407S}).}. There are two additional branches of the O-mode, 
$lt_1$-mode and $lt_2$-mode, that are mixed longitudinal-transverse in nature. 
The $lt_1$-mode is sub-luminal and when propagating along the magnetic field it 
is an arbitrarily polarized purely transverse wave, while during oblique 
propagation it has longitudinal-transverse character. The $lt_2$-mode during 
oblique propagation has longitudinal-transverse nature and is super-luminal for 
low wave numbers but can transform into sub-luminal waves at high frequencies. 
When propagating along the magnetic fields the $lt_2$-mode coincides with the 
electrostatic Langmuir mode. 

There is growing evidence that the PPAs of high linearly polarized time samples
in pulsars follow the RVM, similar to PSR J0332+5434, with two orthogonal 
tracks. CCR is the only known mechanism that can excite waves polarized 
parallel and perpendicular to the curved magnetic field line planes such that
the polarization vector of the emission can exhibit orthogonal RVM 
characteristics \citep{2009ApJ...696L.141M,2023MNRAS.521L..34M,
2023ApJ...952..151M}. As a result the observed PPM and the SPM of the highly 
polarized signals can be readily associated with the X (or $t$-mode) and O (or 
$lt$-mode) modes in the strongly magnetized pair plasma, polarized either 
perpendicular or parallel to the curved magnetic field line planes, respectively. The 
association of the PPA with the X-mode has been established in the radio 
emission from the Vela pulsar. The X-ray observations was used to determine its
projected rotation axis in the sky plane while the SG point of the PPA traverse
provided the location of the dipolar magnetic fiducial plane. The difference 
between the projected angle of the rotation axis and the PPA at the SG point is
$\sim 90\degr$, suggesting that the emerging polarization is perpendicular to 
the dipolar magnetic fiducial plane, i.e. like the X-mode 
\citep{2001ApJ...556..380H}. Only one PPA track is observed in the Vela pulsar 
such that the O-mode is absent. Unfortunately, it has not been possible to 
determine the direction of the rotation axis in other pulsars, hindering direct
measurement of the orientation of the emerging radiation. However, the proper 
motion (PM) of Vela pulsar is along its rotation axis, and if this is a general
trend in pulsars then the difference between the PM and the PPA at the SG point
can be used as a substitute to find the direction of the emerging radiation. In
PSR J0332+5434, \cite{2007MNRAS.379..932M} estimated $\mid$PM-PPA$\mid$ for the 
PPM to be $\sim 100\degr$, which is nearly orthogonal to the magnetic fiducial 
plane, and suggested the PPM to be associated with the X-mode, and as a 
corollary the SPM can be identified as the O-mode. However, such associations 
need to be treated with caution, since the orientation of PM to be along the 
rotation axis is not an established fact in the general pulsar population. PSR
J0332+5434 is a relatively old pulsar and the Galactic potential may have 
changed the direction its PM away from the natal kick along the rotation axis 
(\citealt{2013MNRAS.430.2281N}). 
However, if we ascribe the PPM to the X-mode then the spiky emission is a 
feature of the X-mode. The broad highly polarized features exhibit PPM as well
as SPM and hence can associated with both the X-mode and O-mode.

The high linearly polarized spiky emission are correlated across the entire
frequency range and do not show any signature of RFM. This suggests that the 
CCR leading to these spiky features must arise from a narrow range of emission 
heights. A suitable candidate for charge bunches giving rise to CCR is the 
stable relativistic charged envelopes or solitons that can form in the pulsar 
plasma (\citealt{2000ApJ...544.1081M,2018MNRAS.480.4526L,2022MNRAS.516.3715R}). 
The two stream instability develops in the non-stationary plasma and 
facilitates the linear growth of Langmuir waves. The modulational instability 
of the Langmuir waves leads to the formation of charged solitons. The charged 
solitons move along curved magnetic field lines and emit curvature radiation to
excite the $t$ and $lt_1$ modes. In the presence of strong magnetic field the 
refractive indices in the plasma are different for $t$ and $lt_1$ modes, 
resulting in different phase velocities. The $t$-mode (X-mode) has vacuum like 
character and can escape readily from the plasma. The $lt_1$ mode (O-mode) gets 
ducted along the magnetic field lines, and can only escape from the plasma 
boundary. The non-stationary plasma flow is generated from sparking discharges 
in an inner acceleration region (IAR) above the polar cap, which is partially 
screened by ions emitted from the stellar surface \citep[see][for description 
of the sparking process]{2022ApJ...936...35B}. The sparking process forms dense
intermittent plasma clouds outside the IAR that flows outward along the 
magnetic field lines, interspersed with less dense regions. At the radio 
emission region a large number of solitons are formed in these dense plasma 
clouds. The radio emission detected by the observer is obtained from  
incoherent addition of a large number of solitons, each emitting CCR.

The curvature radiation spectrum from a single soliton in vacuum, of size $a_s$
and Lorentz factor $\gamma_s$, moving along strong magnetic field, with radius 
of curvature $\rho$, has been estimated by \citet{2000ApJ...544.1081M}. 
The soliton in plasma moves with the group velocity with 
Lorentz factor being slightly larger than the plasma. 
When compared with the single particle spectra the soliton spectra has an 
extra term $[ 1- \cos(a(\omega/\omega_c) ]^2$, where $a=a_s {\gamma_s}^3/\rho$ 
and the characteristic frequency $\omega_c = 1.5 {\gamma_s}^3 (c/ \rho)$.
These differences causes the curvature radiation spectra of solitons to have 
characteristic frequency to be a few times greater than the single 
particle $\omega_c$ and the spectra is wider and more symmetrical (see 
Fig 4. of \citealt{2000ApJ...544.1081M}).
The soliton spectra power falls off on either side of the characteristic frequency 
and has an upper 
cutoff frequency, $\nu_h \sim c/a_s$, below which the emission remains coherent.
In the presence of plasma the emission from soliton is modified in
a few different ways. Firstly, the emission splits into the $t$ and $lt$ modes 
such that the power in the $t$ mode is seven times weaker than the $lt$ mode. 
This reduction of power is uniform across all frequencies, ensuring that the
spectral shape remains unchanged. Secondly, the emission can get suppressed 
within the plasma. \citet{2004ApJ...600..872G} showed that the $t$-mode is largely 
suppressed compared to the vacuum case, however can still escape freely 
preserving its inherent vacuum like
polarization properties and spectral shape and can also explain the observed 
pulsar luminosities. The $lt$ mode can get entirely 
damped, and requires special regions such as the plasma edge at the boundary 
between clouds to escape (\citealt{2014ApJ...794..105M,2023ApJ...952..151M}). 
Next, if the distance between the solitons in the 
plasma is $D_s$, there is a lower frequency cutoff in the spectrum $\sim 
c/D_s$. Finally, the resultant spectrum is formed after incoherent addition of
emission from a large number of solitons and can deviate from a single soliton 
spectral shape due to variations in several parameters such as $a_s$, 
$\gamma_s$, $\rho$, $D_s$, etc (see \citealt{2022ApJ...927..208B}).

The absence of RFM in the spiky polarized features (see Fig.~\ref{fig9}) can be
interpreted as the emission arising from a single plasma cloud due to 
incoherent addition of CCR from a large number of solitons. In many of these 
features the spectra have inverted parabolic shapes, e.g. the two pulses shown 
in Fig.~\ref{fig8}, pulse number 1051 with $\nu_c \sim 600$ MHz and pulse 6004 
with $\nu_c \sim 500$ MHz. At the emission height of $\sim$ 300 km, we estimate 
$\rho \sim 3\times 10^8$cm in the conal region, which implies $\gamma_s \sim$
300 for these characteristic frequencies. This shows
that spectral shapes with different characteristic frequencies can be obtained
from small variations in $\gamma_s$. However, estimating the other parameters 
like $a_s$ and $D_s$ requires measurements over a wider frequency range. The 
broader polarized features (see Fig.~\ref{fig10} and \ref{fig11}) on the 
contrary show RFM, indicating that the different frequencies arise from a range
of heights. In this case several clouds from different heights contribute to 
the emission at any given frequency in a manner consistent with RFM in the 
average feature. The resultant spectrum is composed of emission from different 
clouds, where each cloud is characterized by different estimates of the 
parameters $a_s$, $\gamma_s$ and $D_s$. Thus the observed spectra reflects an 
average of the distribution of the parameters, and cannot uniquely represent a 
single soliton spectrum. A proper modeling of the spectra requires detailed
simulations which is beyond the scope of this work, and will be addressed in 
future studies.

\section{Summary}\label{sec5}
In this work we have used observations of PSR J0332+5434 over a wide frequency 
range, between 300 MHz and 750 MHz, to study the behaviour of high linearly 
polarized emission features in the single pulses as a function of frequency. 
The time samples with high levels of linear polarization across the entire 
frequency range have their PPAs distributed along two parallel tracks that 
follow the RVM. The two tracks, divided into PPM and SPM, can be associated 
with X-mode and O-mode in a strongly magnetized plasma. Using predictions of 
the A/R effect, the location of the radio emission region in PSR J0332+5434 at 
the observing frequencies is found to be few hundred km above the neutron star 
surface. At the location of radio emission, the polarization behaviour strongly
favours the CCR from charged bunches as the radio emission mechanism. In the 
strongly magnetized, relativistic pair plasma in the pulsar magnetosphere, with
a non-stationary flow along the open field lines, charged solitons are expected
to form, and are the likely candidates for the charged bunches producing CCR.

We find several examples of high linearly polarized narrow spiky emission 
features whose location within the profile is roughly constant across the 
frequency range. This emission is always associated with the PPM, and can also
possibly be associated with the X-mode. We suggest that this spiky emission
arises from a single plasma cloud with incoherent addition from a large number 
of solitons, each emitting CCR. We detected many cases where the spectra of the
spiky emission features have inverted parabolic shapes, that resemble the 
soliton spectrum. This further implies that in an emission cloud there is 
little spread in the values of soliton size, $a_s$, inter soliton distance, 
$D_s$ and Lorentz factor of the solitons, $\gamma_s$. The peak of the inverted 
spectrum can be used to find $\gamma_s$ which is of the order of a few hundred.
There are also examples of high linearly polarized emission features with much 
broader size. These features seem to follow the RFM seen in the average 
profile, and it can be interpreted that their emission arise from a range of 
heights. The spectra of these features are formed after averaging over many 
clouds that have different distribution of soliton parameters. This provides a 
qualitative description of the origin of RFM in pulsars, although more detailed
quantitative estimates are necessary to understand how the averaging process 
works in practice.

\begin{acknowledgments}
	 We thank the referee for useful comments.
D.M. acknowledges the support of the Department of Atomic Energy, Government of
India, under project No. 12-R\&D-TFR-5.02-0700. This work was supported by the 
grant 2020/37/B/ST9/02215 of the National Science Center, Poland.
\end{acknowledgments}




\bibliography{References}{}
\bibliographystyle{aasjournal}

\end{document}